\documentclass[prb,twocolumn,superscriptaddress,citeautoscript,showpacs,amsart]{revtex4-1}

\usepackage{graphicx}
\usepackage{multirow}
\usepackage{color}
\usepackage{bm}
\usepackage{times}
\usepackage{amsmath,bm,amsfonts}
\usepackage{dcolumn}
\usepackage{graphicx}
\usepackage{latexsym}
\usepackage{cancel}
\usepackage{hyperref}
\usepackage{physics}

\usepackage{ulem} 
\usepackage{braket}

\def\ket#1{|#1\rangle }
\def\bra#1{\langle #1 |}
\def\n{\nonumber \\ }

\newcommand{\ma}{\sigma}

\newcommand{\dt}{\delta}
\newcommand{\ep}{\epsilon}
\newcommand{\la}{\lambda}
\newcommand{\f}{\frac}
\newcommand{\ta}{\theta}

\newcommand{\pa}{\partial}
\newcommand{\Dt}{\Delta}
\newcommand{\dg}{\dagger}
\newcommand{\kb}{\mathbf{k}}
\newcommand{\Sb}{\mathbf{S}}
\newcommand{\rb}{\mathbf{r}}
\newcommand{\ab}{\mathbf{a}}

\newcommand{\rw}{\rightarrow}

\renewcommand{\thefigure}{\arabic{figure}}
\renewcommand{\thetable}{\arabic{table}}

\renewcommand{\arraystretch}{1.5}
\newcommand{\A}{\alpha}
\newcommand{\B}{\beta}

\newcommand{\w}[1]{\omega_{#1}}
\newcommand{\al}[1]{\langle #1 \rangle}

\newcommand{\Eb}{\mathbf{E}}

\newcommand{\qb}{\mathbf{q}}
\newcommand{\up}{\uparrow}
\newcommand{\dw}{\downarrow}
\newcommand{\nm}{\nonumber \\&}
\newcommand{\alg}[1]{\begin{align}#1\end{align}}
\newcommand{\mtx}[1]{\left(\begin{matrix}#1\end{matrix}\right)}

\newcommand{\de}{\tilde}

\newcommand{\ah}{\hat{a}}
\newcommand{\bh}{\hat{b}}

\newcommand{\nq}{\nonumber \\ =&}
\newcommand{\Dn}{\Dt n}
\newcommand{\Dw}{\Dt w}

\begin{document}

\title{Emergent inductance from spin fluctuations in strongly correlated magnets}

\author{Taekoo \surname{Oh}}
\affiliation{RIKEN Center for Emergent Matter Science (CEMS), Wako, Saitama 351-0198, Japan}

\author{Naoto \surname{Nagaosa}}
\email{nagaosa@riken.jp}
\affiliation{RIKEN Center for Emergent Matter Science (CEMS), Wako, Saitama 351-0198, Japan}

\date{\today}

\begin{abstract}
Recently, the intriguing phenomenon of emergent inductance has been theoretically proposed and
experimentally observed in nanoscale spiral spin systems subjected to oscillating currents.
Building upon these recent developments, we put forward the concept of emergent inductance in 
strongly correlated magnets in the normal state with spin fluctuations.
It is argued that the inductance shows a positive peak at temperatures above  the ordering temperature. 
As for the frequency dependence, in systems featuring a single-band structure or a gapped multi-band, 
we observe a Drude-type, while in gapless multi-band systems, a non-Drude inductance with a sharp 
dip near zero frequency. 
These results offer valuable insights into the behavior of strongly correlated magnets and open up 
new possibilities for harnessing emergent inductance in practical applications.
\end{abstract}


\maketitle


{\it Introduction.|}
The noncollinear magnets show a variety of intriguing phenomena such as multiferroics of spin origin~\cite{sergienko2006role,bezvershenko2018stabilization}, topological protection of spin textures~\cite{schulz2012emergent,hirschberger2019skyrmion}, and various kinds of Hall effects~\cite{taguchi2001spin,machida2007unconventional,neubauer2009topological,nagaosa2013topological,kim2020strain,li2021correlated,song2022higher}. 
The underlying principle of these phenomena is the emergent electromagnetic field (magnetic ($\bm{h}$) and electric ($\bm{e}$) fields)~\cite{barnes2007generalization} associated with the spin Berry connection $a_\mu$ ($\mu = 0 (t), i= 1,2,3$)
defined by the spin direction field 
$\bm{n}(r,t)$ as  
\begin{align}
h_i= (\bm{\nabla}\times \bm{a})_i=
\frac{\hbar c}{2e}\left(
\varepsilon_{ijk} \bm{n}\cdot
\partial _{j}\bm{n}\times \partial _{k} \bm{n} \right), 
\label{eq:emf}
\end{align}
where $\varepsilon_{ijk}$ is the totally antisymmetric tensor, and 
\begin{align}
e_i = \frac{1}{c} \frac{\partial a_0}{\partial x_{i}} 
-\frac{\partial a_i}{\partial t}=
\frac{\hbar c}{2e}\left(
\bm{n}\cdot
\partial_{i}\bm{n}\times \partial_{t}\bm{n}\right).
\label{eq:eme}
\end{align}  
Applying these formula to the spiral magnet, the emergent inductance has been theoretically proposed~\cite{nagaosa2019emergent}, and experimentally demonstrated in Gd$_3$Ru$_4$Al$_{12}$~\cite{yokouchi2020emergent}. 
Later, the emergent inductance has been observed in YMn$_6$Sn$_6$ beyond room temperature~\cite{kitaori2021emergent}.
Emergent inductance in Rashba spin-orbit-coupled system has been theoretically proposed~\cite{ieda2021intrinsic}.

In respect to the applications, the frequency dependence and the quality factor $Q$ are important issues. 
At present, the experiments show the rapid decay of inductance as the frequency increases above $\sim 10$kHz, and the quality factor $Q$ is less than a few $\%$.  
This characteristic frequency is considered to be due to the collective dynamics of the ordered spin system.
Typically, the dynamics is characterized by the energy scale $\alpha J$, with the Gilbert damping 
constant $\alpha~ ( \sim 0.01)$ and the exchange coupling $J$ of the order of 1~meV. 
It is note-worthy that the energy scale corresponds to the order of 1~GHz, which is much larger than 
the observed one but much smaller than conduction electrons' that is typically of the order of 1~THz.

Therefore, in the present paper, we propose a novel way to improve the frequency dependence of the 
emergent inductance by utilizing the rapid quantum/thermal spin fluctuation with higher energy than 
that of ordered moments.
To explore this phenomenon, we employ the U(1) Slave-fermion theory, where the spin Berry connection $\bm{a}$ 
appears naturally as the phase of the singlet correlation of neighboring spins, which can also be interpreted 
as the gauge potential.
In this formalism, the electrons undergo fractionalization into distinct entities known as spinons and holons.
Importantly, the spinons possess significantly longer lifetimes compared to the holons, leading us to anticipate 
the emergence of spinonic inductance in the low-frequency regime. Remarkably, the spinon inductance is 
physically observable according to Ioffe-Larkin composition rule \cite{ioffe1989gapless}, 
which is implied by the coupling between spinons and holons through the gauge field $\bm{a}$ 
in Schwinger boson method \cite{PhysRevB.38.316} or $U(1)$ Slave-fermion theory \cite{RevModPhys.78.17}. 

\begin{figure}[t]
    \centering
    \includegraphics[width=\columnwidth]{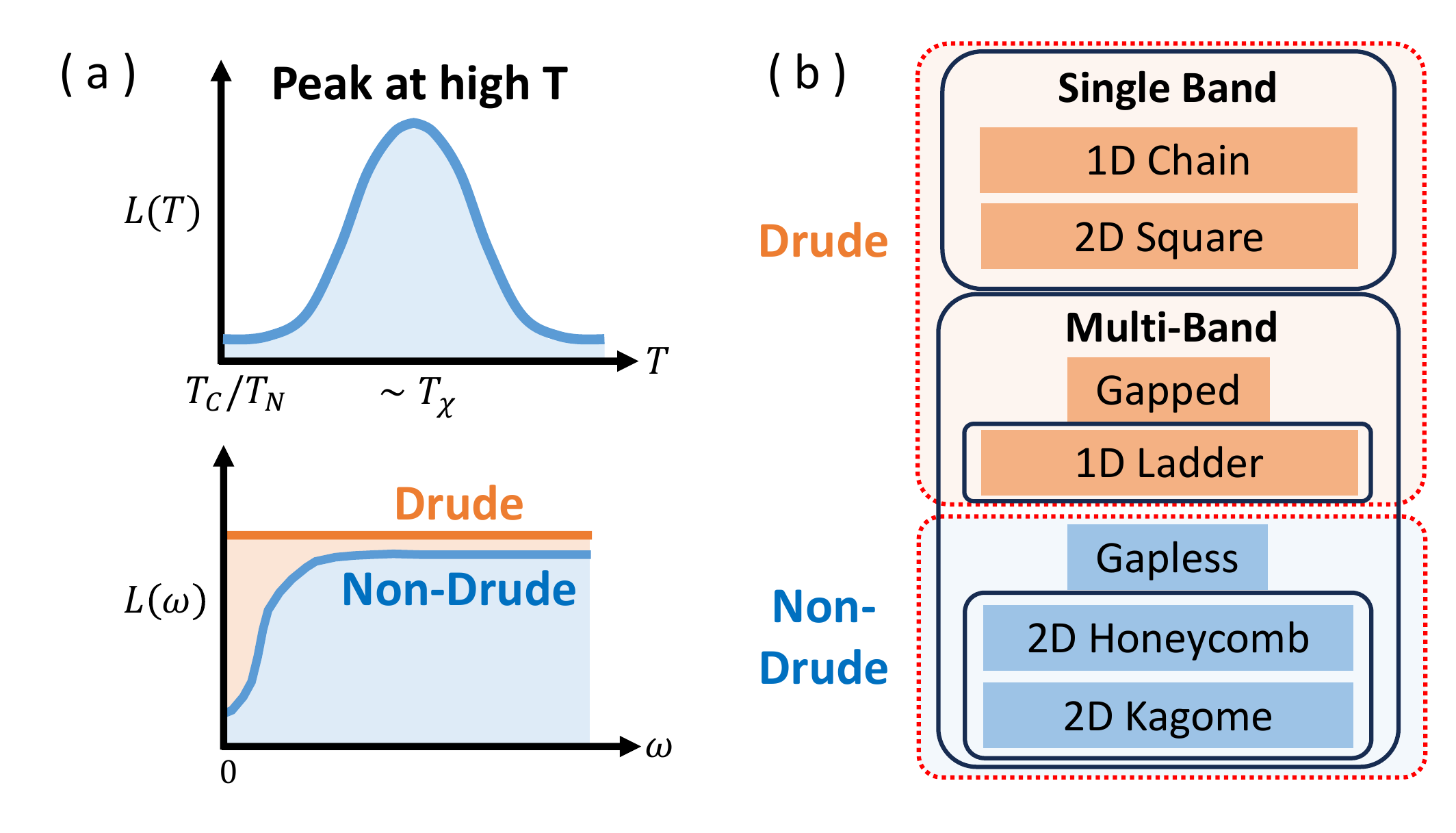}
    \caption{Our main findings of emergent inductance in strongly correlated magnets at high temperatures.
    (a) The schematics of inductance in temperatures (upper panel) and in frequencies (lower panel). 
    (b) The classification of inductance: Drude and non-Drude types.}
    \label{fig:1}
\end{figure}

Our findings, which is illustrated by several systems including 1D spin chain, 1D spin ladder, 2D square, 2D honeycomb, 
and 2D Kagome lattices, can be summarized as follows.
First, we observe that the inductance displays a positive peak at temperatures higher than 
the "ordering temperature." [See the upper panel of Fig.~\ref{fig:1}(a).]
Note that the ordering temperature here means the characteristic temperature where the correlation length grows rapidly since there is no long range ordering in 1D and 2D Heisenberg models at finite temperatures.
This behavior is attributed to the increased resistivity and associated inductance as the system becomes less metallic near the spinon phase transition temperature. 
At much higher temperature, as the system shows the thermally assisted hopping conduction, 
we anticipate that the inductance lowers down and the peak is found.
Second, as for frequency dependence, we distinguish between Drude-type inductance in single-band or 
gapped multi-band systems and non-Drude-type inductance in gapless multi-band systems. [See Fig.~\ref{fig:1}(b).] 
Drude-type inductance remains independent of frequency, while non-Drude-type inductance exhibits 
a sharp negative dip near $\w{}=0$, as depicted in the lower panel of Fig.~\ref{fig:1}(a).


{\it Method |} 
To investigate the phenomenon of inductance in strongly correlated magnets, 
we employ the Slave-fermion method~\cite{barnes1976new,coleman1984new, lee1992gauge,lee2006doping}.
The behavior of strongly correlated magnets can be effectively captured by the widely 
studied $t$-$J$ model, defined by the 
Hamiltonian:
\begin{align}
    H = - \sum_{\al{ij}} J(\Sb_i\cdot\Sb_j - \f{1}{4}n_in_j) - \sum_{\al{ij}}t_{ij}( c_{i\A}^\dg c_{j\A} + \text{h.c.}).
\end{align}
Within the Slave-fermion method, the electron operator can be expressed as 
$c_{i\alpha}^\dagger = f_{i\alpha}^\dagger b_i + \epsilon_{\alpha\beta} f_{i\beta} d_i^\dagger$, subject to the constraint 
$\sum_{\alpha} f_{i\alpha}^\dagger f_{i\alpha} + b_i^\dagger b_i + d_i^\dagger d_i = 1$. 
Here, $b_i^\dagger$ (holon) represents the vacancy, $f_{i\alpha}^\dg$ (spinon) denotes the single-electron state 
with spin $\alpha$, and $d_i^\dagger$ (doublon) corresponds to the double-occupancy.
Since the electron is a fermion, either $f_{i\alpha}$ or $b_i/d_i$ must be fermionic, while the other is bosonic. 
Specifically, when $b_i/d_i$ exhibits fermionic (bosonic) behavior, it is referred to as the Slave-fermion (Slave-boson) theory. 

Due to strong correlation effects, the presence of double occupancy is prohibited. 
Therefore, the electron operators can be expressed as $c_{i\A}^\dg = f_{i\A}^\dg b_i$, subject to the constraint 
$\sum_\A f_{i\A}^\dg f_{i\A} + b_i^\dg b_i = 1$. 
Introducing the operators $\hat \chi_{ij} = \sum_{\A} f_{i\A}f_{j\A}$ and $\hat \Delta_{ij} = \sum_{\A\B}\epsilon_{\A\B}f_{i\A}f_{j\B}$, we find that $\mathbf{S}_i \cdot \mathbf{S}_j = \f{1}{2}(\hat \chi_{ij}^\dg \hat \chi_{ij}- 2S(S+1) )$ for ferromagnets, while $\mathbf{S}_i \cdot \mathbf{S}_j = \f{1}{2}( 2S^2 -\hat \Dt_{ij}^\dg \hat \Dt_{ij})$ for antiferromagnets. $\hat \chi_{ij}$ 
represents coherent spinon propagation and $\hat \Delta_{ij}$ represents spinon singlet-coupling.

By employing Slave-fermion mean-field theory (SFMFT) and introducing the order parameters 
$\chi_{ij} = \langle \hat \chi_{ij} \rangle$ in ferromagnets, we arrive at the following Hamiltonian:
\begin{align}
H =& -\tilde J\sum_{\langle ij \rangle} (\chi_{ij} \hat \chi_{ij}^\dg + \chi_{ij}^* \hat \chi_{ij}) + \sum_{\langle ij \rangle} 
(t_{ij} \chi_{ij} b_i^\dg b_j + \text{h.c.}) \n
&+ \sum_i \lambda_i (\sum_\A f_{i\A}^\dg f_{i\A} + b_i^\dg b_i - 1).
\end{align}
Here $\tilde J = J/2$. It should be noted that the last line represents the Lagrange multiplier $\la_i$ 
associated with the constraint, and the fermionic and bosonic components are treated separately in this formulation.


\begin{figure}
    \centering
    \includegraphics[width=\columnwidth]{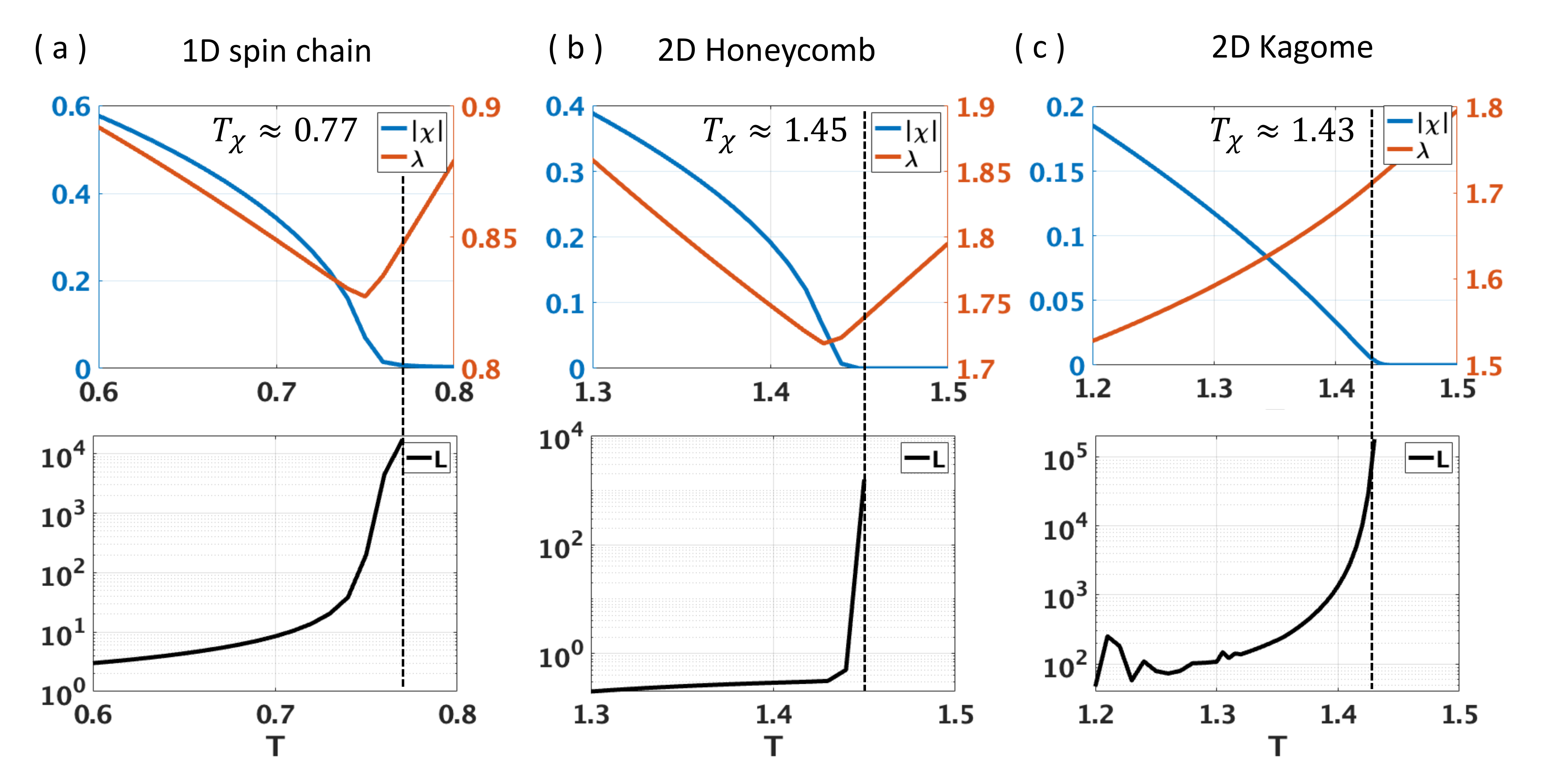}
    \caption{The order parameters (upper panel) and log-scale inductance (lower panel) in temperatures from the 
self-consistent SFMFT with $\tilde J = 1$ on (a) 1D spin chain, (b) 2D honeycomb lattice, and (c) 2D Kagome lattice. 
$T$ is in units of $\tilde{J}$, and $L$ is in units of $\sim 0.1~\mu$H for the 100~nm cubic sample.} The blue lines are the coherent spinon propagation $\chi$, the orange lines are the Lagrange multiplier $\la$, the black lines are the maximum of inductance $L$ in $\w{} \in [-3,3]$, and the dotted lines denote $T_\chi$.
    \label{fig:2}
\end{figure}


Using a $U(1)$ gauge theory, the physical conductivity is determined by $\ma^{-1} = \ma_f^{-1} + \ma_b^{-1}$, where $\ma_{f,b}$ represent the conductivity of spinons and holons, respectively. 
This is known as the Ioffe-Larkin composition rule~\cite{ioffe1989gapless}. The rule arises from the fact that the spinons flows against holons because of the strong coupling of holons and spinons by gauge field.
We assume that the system is appreciably away from half-filling, so the holon conductivity $\ma_b$ follows a 
Drude-type behavior, $\ma_b = \ma_0/(1 - i\w{}\tau_b)$, which is relatively temperature-insensitive
and much larger than that of spinons. Here, the transport lifetime $\tau_b$ is typically $\sim 1$~THz$^{-1} = 1~$ps,
and the contribution to the inductance from the holons can be also neglected.
Therefore, the remaining part of the Hamiltonian describes the Schwinger-boson theory for the spinons. 

Spinons exhibit conductivity at half-filling only in ferromagnetic systems ($J>0$) since $\chi = \chi_{ij}=0$ in antiferromagnetic systems ($J<0$). Thus, we consider the ferromagnetic model on various lattices, such as spin chains, honeycomb, and Kagome lattices, and determine the order parameters 
self-consistently at temperatures, as shown in the upper panels of Fig.~\ref{fig:2}. 
Subsequently, we compute the current-current correlation function $\Pi(\qb,\tau) = -\al{T_{\tau} J(\qb,\tau) J(-\qb)}$ 
and obtain $\text{Re}~\ma(\w{}) = -\text{Im}~\Pi(\w{})/\w{}$ by analytic continuation $i\w{n} \rw \w{}+i\eta$. 
We mostly set $\qb=0$ as the external electric field is constant in space while oscillating in time.
The imaginary conductivity is evaluated using the Kramers-Kronig relation, 
$\text{Im}~\ma(\w{}) = -\frac{1}{\pi} \int d\w{}' \frac{\text{Re}~\ma(\w{}')}{\w{}'-\w{}}$. 
The inductivity is then calculated as $\mathcal{L} = -\text{Im}~\rho(\w{})/\w{}$, where $\rho(\w{}) = 1/\ma(\w{})$.
The inductance can be obtained by $L=\mathcal{L}l/A$ where $l$ is the length and $A$ is the area of the system. 
We set $\tilde J=1$, and present energies and frequencies in units of $\tilde J$. The frequency range is $\w{} \in [-3,3]$ (up to $\sim 1$ GHz), and the spinon lifetime parameter is $\eta \approx dk$ where $dk$ is the $k$-mesh size.
We suppose that the system is a cube with $100$~nm in all Figures, so the unit of $L$ is $\sim0.1~\mu$H.
We will discuss about how the units are determined later.
Further computational details can be found in the Supplementary Materials (SM).


{\it The inductance peak at high T |}
In the upper panels of Fig.~\ref{fig:2}, the order parameters $\chi$ and $\lambda$ are plotted as functions of temperatures.
The phase transition to the finite $\chi$ occurs at $T_\chi \approx 0.77$ for the 1D spin chain, 
$T_\chi \approx 1.45$ for the 2D honeycomb lattice, and $T_\chi \approx 1.43$ for the 2D Kagome lattice. 
The inductance $L$ can only be determined below $T_\chi$ by SFMFT, since $\chi$ is finite only for this regime.
This phase transition is an artifact of the mean-field theory, and describes the crossover from the coherent propagation of the spinons to their thermal hopping conduction.
This limitation arises because the model introduces artificial behavior where $\chi$ approaches zero at high temperatures and fails to capture the short-range spin correlations which persist at finite values at any finite temperatures~\cite{lee2006doping}.
It is also noted here that the finite energy gap $E_g$ of the lowest spinon dispersion similar to or smaller than temperatures below $T_{\chi}$. [See SM.]

In the lower panels of Fig.~\ref{fig:2}, the increased inductance near $T_{\chi}$ within the frequency range $\omega \in [-3,3]$ is shown.
The inductance reaches its highest value $L = 10^3 \sim 10^5$ near $T_\chi$ for every case.
Despite the anticipated exaggeration of values from the artifact of SFMFT, the increment tendency of inductance remains true near $T_{\chi}$.
It is also important to note that $T_\chi > T_C$ (the Curie-Weiss temperature) considering the presence of short-range spin correlation near $T_{\chi}$. 

On the other hand, beyond $T_{\chi}$ where SFMFT fails, we anticipate that the conductivity rises again and the inductance decreases as the thermally activated hopping motion increases the conductivity with temperatures. 
At higher temperatures where lattice vibrations are more significant, the transfer of spinons could be primarily governed by incoherent thermal excitations. 
This excitations come from the electron-phonon coupling and consequent polaron effect which is not included in the present model~\cite{HOLSTEIN1959325,fetherolf2020unification}. 
It is argued that the conductivity has the minimum at the crossover from the coherent propagation to the hopping conduction associated by the phonon. 
Therefore, the positive peak of the inductance is expected near $T_{\chi}$. 


{\it The inductance in frequencies|} 
The Drude-type inductance can be briefly reviewed as follows~\cite{resta2018drude}. 
In a normal metal subjected to an AC electric field $\Eb(\w{})$, the Drude conductivity is 
given by $\ma = \ma_0/(1-i\w{}\tau)$, where $\ma_0 = ne^2\tau/m_e$. 
Here, $n$ represents the number density, $e$ is the charge of the carriers, $\tau$ is the relaxation time, 
and $m_e$ is the mass of the carriers. 
Consequently, the resistivity is given by $\rho = (1-i\w{}\tau)/\ma_0$, and the associated inductance is given 
by $L =\tau l/\ma_0 A = m_e l/ne^2 A$.
Importantly, the inductance $L$ remains frequency-independent. 
In the present case, the thermally activated spinons across the gap contributes to the Drude-like transport due to $E_g \lesssim T$.

In the upper panels of Fig.~\ref{fig:3}, we show the energy band structures near $T_\chi$ for the spin chain, honeycomb lattice, and Kagome lattice. 
The 1D spin chain exhibits a single-band structure.
In contrast, the 2D honeycomb and Kagome lattices are multi-band systems with band crossing points, where the inter-band contribution to the conductivity is also finite.
In the honeycomb lattice, the band crossing occurs at the $K$ points. 
For the Kagome lattice, band crossings are observed at both the $\Gamma$ and $K$ points.


\begin{figure}[t]
    \centering
    \includegraphics[width=\columnwidth]{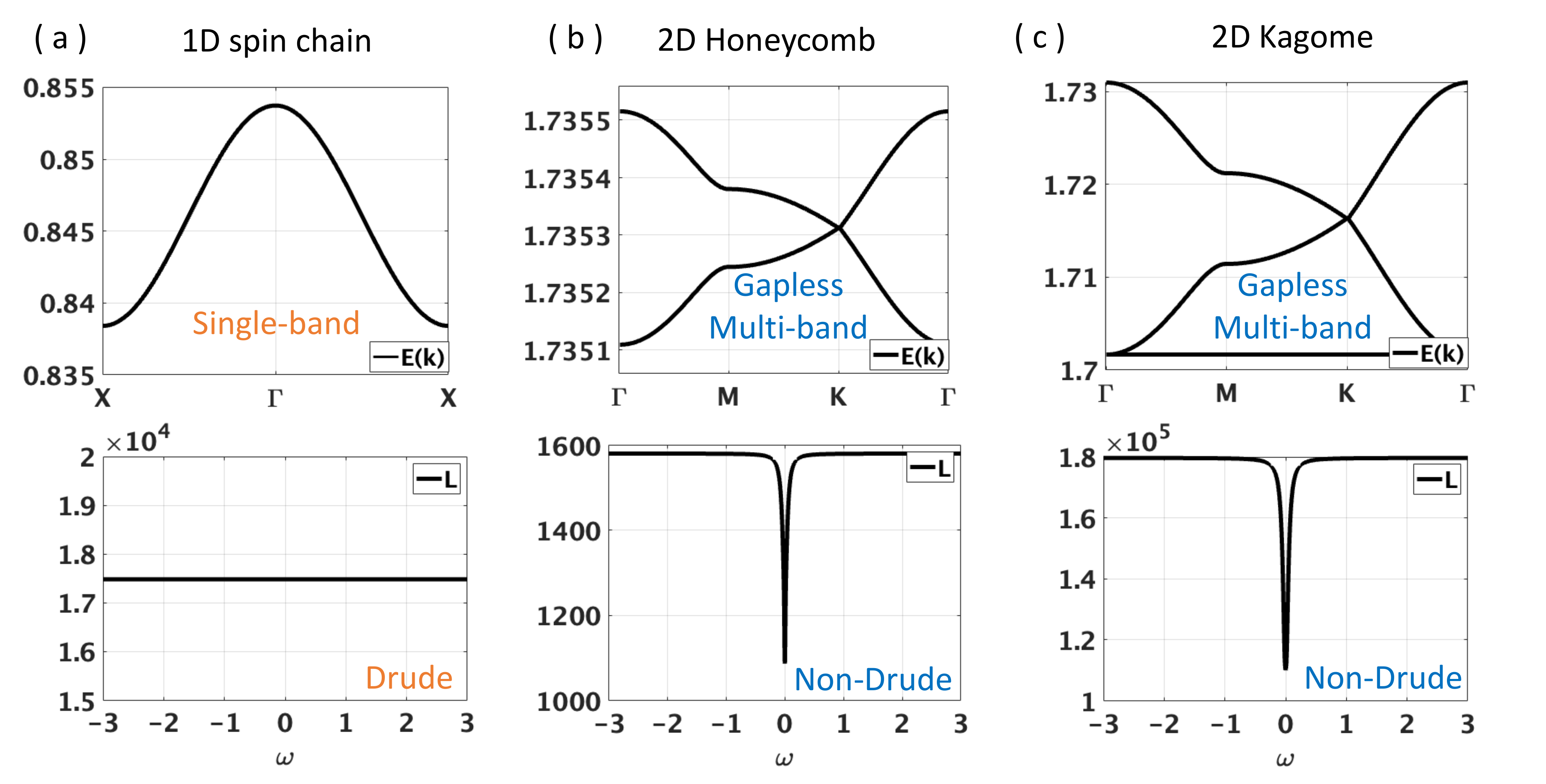}
    \caption{The energy band (upper panel) and inductance in frequencies (lower panel) near $T_\chi$, for (a) 1D spin chain, (b) 2D honeycomb lattice, and (c) 2D Kagome lattice. $E(k)$ and $\w{}$ are in units of $\tilde J$, and $L$ has the same unit as above. The 1D spin chain is a single-band system, exhibiting Drude-type inductance. The honeycomb and Kagome lattices are gapless multi-band systems, exhibiting a sharp negative dip structure near $\w{}=0$.
    }
    \label{fig:3}
\end{figure}


In the lower panels of Fig.~\ref{fig:3}, we present the numerically computed inductance $L$ near $T_\chi$ for the spin chain, 
honeycomb lattice, and Kagome lattice. 
Notably, the inductance exhibits frequency independence only for the spin chain, following a Drude-type behavior. 
However, for the honeycomb and Kagome lattices, the inductance displays a sharp negative dip structure near $\w{} = 0$.
The characteristics of this sharp dip, including its depth and width, are closely connected to the 
lifetime parameter $\eta$ of the spinons. 
This indicates that the resonance structure is rooted in the transport phenomena of these systems. 
The energy gap $E_g \lesssim T$ does not appear in the frequency dependence of the inductance because of the thermally activated bosons.
Further details can be found in the SM.

The distinction between the spin chain and other systems arises from the interband transitions at $\qb = 0$ in spinon transport phenomena. 
This can be demonstrated through analytic calculations of inductance in two scenarios.

First, consider the 1D spin chain. The current-current correlation function is given by
\begin{align}
\Pi^1(\qb,\w{}) = 2(\tilde J\chi)^2 \int \frac{dk}{2\pi} \sin^2 k 
\frac{-\Delta \omega_{k,q} \Delta n}{(\w{}+i\eta)^2 - \Delta \omega_{k,q}^2},
\end{align}
where $\Delta \omega_{k,q} = \omega_{k-q/2} - \omega_{k+q/2}$, 
$\Delta n = n(\omega_{k-q/2}) - n(\omega_{k+q/2})$, and $\omega_{k} = \lambda - \tilde J|\chi| \cos k$ represents the energy. 
Here, $n$ denotes the Bose-Einstein distribution. 
Notably, the intraband transition described by $\Delta \omega_{k,q}$ dominates because of a single band in the system.
By taking the limits $\eta \to 0$ and $\qb \to 0$ successively, the total conductivity becomes
\begin{align}
\ma^1(\w{}) = -2(\tilde J\chi)^2 \int \frac{dk}{2\pi} \sin^2 k \left(\pi \delta(\w{}) + \frac{i}{\w{}}\right)n'(\omega_{k}).
\end{align}
The integration in $k$ yields $\ma^1(\qb,\w{}) = A(\pi \delta(\w{}) + i/\w{})$ with $A>0$, rendering the inductance 
$L = 1/A$ frequency-independent.

Second, consider the 2D honeycomb lattice. The current-current correlation function at $\qb = 0$ is given by
\begin{align}
\Pi^2(0,\w{}) = 4(\tilde J\chi)^2 \int \frac{d^2k}{(2\pi)^2} \frac{\Delta n}{\w{k}} \frac{g(\kb)^2}{(\w{}+i\eta)^2 - (2\w{k})^2},
\end{align}
where $\kb = (k_1,k_2)$, $g(\kb) = (\tilde J\chi) [1+\cos k_1 + \cos(k_1-k_2)]$, $\w{k}^2 
= (\tilde J \chi)^2[3+2\cos k_1+2\cos k_2 + 2\cos(k_1-k_2)]$, and $\Delta n = n(\lambda - \w{k}) - n(\lambda + \w{k})$. 
Remarkably, the dispersion of the energy band in the honeycomb lattice is 
$\lambda \pm \w{k}$, so $2\w{k} = (\lambda + \w{k}) - (\lambda - \w{k})$ corresponds to interband transitions. 
Upon performing some algebraic manipulations, we arrive at
\begin{align}
\ma^2(\w{}) = 2(\tilde J\chi)^2 \int_k \frac{ g(k)^2\Delta n}{\w{k}^3} 
\frac{i(\w{}+i\eta)}{(\w{}+i\eta)^2-(2\w{k})^2}, \label{eq:6}
\end{align}
where $\int_k = \int d^2k/(2\pi)^2$. 
The integrand exhibits resonance behavior observed in numerical computations. 
If a band crossing point exists such that $\w{k} = 0$, in the vicinity of $\w{} = 0$, 
the denominator of the integrand approaches zero, leading to a significant increase in conductivity. 
Consequently, the resistivity decreases near $\w{} = 0$, resulting in a sharp negative 
dip structure in the corresponding inductance.

Three additional aspects are worth noting. 
First, the result obtained for the 1D spin chain can be generalized to higher dimensions with a single band. 
Consequently, we anticipate that higher-dimensional single-band systems would exhibit Drude-type inductance. 
We briefly address that the 2D square lattice hosts Drude-type inductance in SM.
Second, in a multi-band system, the sharp dip structure near $\w{}=0$ may not exist when there is a gap in the energy bands, considering that the denominator in Eq.~\ref{eq:6} would not approach zero for such a case. 
We illustrate it further with a 1D spin ladder model in SM.
Lastly, in the presence of impurities or disorder, even single-band systems or gapped multi-band systems demonstrate non-Drude inductance. 
This arises from the contribution of intraband transitions to the transport phenomenon. 
We utilize the Mattis-Bardeen scheme~\cite{mattis1958theory} in the spin chain, and provide the detail in SM.


{\it Discussion|} 
In summary, we have examined the emergence of inductance in strongly correlated magnets with fractionalized spins. 
At temperatures above the ordering temperature, the dispersion of spinons decreases, leading to a significant 
increase in inductance. 
The type of inductance, whether Drude or non-Drude, depends on the system's characteristics such 
as the number of bands and the presence of band gaps. 
In non-Drude cases, a sharp dip structure near $\w{}=0$ is observed, the width of which is determined 
by the spinon's relaxation rate, which is typically $\sim J \alpha$ with $\alpha$ being the Gilbert damping constant,
and is much smaller than the usual transport relaxation rate $\tau_b^{-1}$.

Here, we discuss about the units of physical quantities and the range of the estimated inductance in our theory. We assume the unit of exchange interaction $J$ is the order of $\sim 1$ meV and the unit of lattice constant $a = 1~\AA$.
Thus, considering that $\hbar = e = 1$, the unit of $k_BT$ is $11.6$ K, that of frequency $\w{}$ is $242$ MHz, that of the spinon lifetime $\eta^{-1}$ 
is $4.13 ~ ns$, that of resistivity $\rho$ is $258 ~\mu \Omega \cdot$ cm, and that of inductivity $\mathcal{L}$ is $1.07$ pH$\cdot$cm. 
Accordingly, in a cubic system with $100~$nm edges, the unit of inductance is approximated $\sim 0.1 \mu$H. Near $T_{\chi}$, $L \sim 10^2 - 10^4 ~\mu$H, which can be contrast to the previous experimental findings show $L \sim 1 - 10~\mu$H~\cite{kitaori2021emergent}. 
Although the SFMFT exaggerates the computed inductance, we predict that the positive inductance peak at high temperatures is experimentally observable in the strongly correlated magnets. 

Lastly, it is important to distinguish our work from a previous theoretical study~\cite{kurebayashi2021electromagnetic}. 
Their investigation focused on the emergence of inductance in spiral magnetic orders with weakly correlated 
electrons and a strong Ruderman-Kittel-Kasuya-Yosida (RKKY) interaction. 
In this case, the spin density wave drives the emergent inductance.
We believe that our results provide useful intuitions into the transport in strongly correlated system complementary to Ref.~\cite{kurebayashi2021electromagnetic} and reveal the 
practical prospects for utilizing emergent inductance.

\section*{Acknowledgment}
This work was supported by JST, CREST Grant Number JPMJCR1874, Japan.

\clearpage

\section*{Appendices to "Emergent inductance from spin fluctuations in strongly correlated magnets"}

\appendix
\tableofcontents
\renewcommand{\thefigure}{S\arabic{figure}}
\renewcommand{\thetable}{S\arabic{table}}
\renewcommand{\arraystretch}{1.5}
\setcounter{figure}{0}
\setcounter{equation}{0}
\setlength{\tabcolsep}{3pt}

\section{Ioffe-Larkin Rule in U(1) Gauge Theory of $t-J$ model }

The strongly correlated system is well described by so-called $t-J$ model,
\begin{align}
	H = \sum_{\al{ij}} J(\Sb_i\cdot\Sb_j - \f{1}{4}n_in_j) - \sum_{ij,\A}t_{ij}( c_{i\A}^\dg c_{i\A} + h.c.), \label{eq:01}
\end{align}
where $t_{ij}$ represents the hopping between $i$ and $j$ sites, and $J$ is the exchange strength. $J<0$ ($J>0$) represents ferromagnetic (antiferromagnetic) exchange. Notably, in the manuscript, we mainly consider $J<0$ system. In this section, we review the Ioffe-Larkin rule of $U(1)$ gauge theory of $t$-$J$ model.

Due to the strong Coulomb repulsion in the system, we impose the constraint that the double occupancy at a site is forbidden, $\sum_{\A} c_{i\A}^\dg c_{i\A} \leq 1$. 
Then, the Slave-boson or Slave-fermion method can be applied here. 
The operator can be replaced by
\begin{align}
	c_{i\A}^\dg = f_{i\A}^\dg  b_i + \ep_{\A\B} f_{i\B} d_i^\dg.
\end{align}
Here, $\ep_{\A\B}$ is the anti-symmetric tensor, $b_i^\dg$ and $b_i$ are the creation / annihilation operators for the 0-electron state (holon) at site $i$, $f_{i\A}^\dg$ and $f_{i\A}$ are the creation / annihilation operators for the 1-electron state with spin $\A$ (spinon) at site i, and $d_i^\dg$ and $d_i$ are the creation / annihilation operators for the 2-electron states (doublon) at site $i$. 
When $c_{i\A}^\dg$ is applied to 0-electron state, the 1-electron state with spin $\A$ is created. When $c_{i\A}^\dg$ is applied to the 1-electron state with spin $\B \neq \A$, it creates 2-electron state. 
Since $c$ is fermionic, one of $b/d$ and $c$ is fermionic and the other is bosonic. 
If $b, d$ are bosons (fermions) and $f$ is the fermion (bosons), we call it "Slave-boson (Slave-fermion) theory." 
A constraint exists for both theories,
\begin{align}
	\sum_\A f_{i\A}^\dg f_{i\A} + b_i^\dg b_i + d_i^\dg d_i = 1.
\end{align}
This means that only the states can only have one of the following states: vacancy, 1-electron with spin $\up$, 1-electron with spin $\dw$, and 2-electron states.
The double occupancy is forbidden by simple delete of $d$ operators.
\begin{align}
	c_{i\A}^\dg = f_{i\A}^\dg b_i.
\end{align}
The constraint changes into
\begin{align}
	\sum_\A f_{i\A}^\dg f_{i\A} + b_i^\dg b_i = 1.
\end{align}

Here, we induce the Ioffe-Larkin rule by applying Slave-boson theory to Eq.~\ref{eq:01}. 
Please note that this argument can be equivalently applied to Slave-fermion theory which we use in the manuscript.
Projecting the Hilbert space onto holon and spinon states, we need to find the correct matrix elements. 
As the spins are represented by $\Sb_i = \f{1}{2}f_{i\A}^\dg \ma_{\A\B} f_{i\B}$, the dot product is represented by
\begin{align}
	\Sb_i\cdot\Sb_j 
	=& -\f{1}{4} ( B_{ij}^\dg B_{ij} + A_{ij}^\dg A_{ij}) +\f{1}{4}f_{i\A}^\dg f_{i\A}.
\end{align}
Here,
\begin{align}
	&A_{ij}^{\dg} = \ep_{\A\B} f_{i\A}^\dg f_{j\B}^\dg, A_{ij} = \ep_{\A\B} f_{j\B}f_{i\A},\n
	&B_{ij}^\dg = f_{i\A}^\dg f_{j\A}, B_{ij} = f_{j\A}^\dg f_{i\A}.
\end{align}
$A_{ij}$ is the spin-singlet coupling, and $B_{ij}$ is the hopping of the spinons.
On the other hand, $n_in_j = (1-b_i^\dg b_i) (1-b_j^\dg b_j)$. Ignoring the hole interaction $b_i^\dg b_i b_j^\dg b_j$, we have
\begin{align}
	H =&- \sum_{\al{ij}} J' (A_{ij}^\dg A_{ij} + B_{ij}^\dg  B_{ij})  + \mu_B \sum_{i}b_i^\dg b_i \nm - \sum_{ij} t_{ij}(f_{i\A}^\dg b_i b_j^\dg f_{j\A}  + h.c.).
\end{align}
Here, $J'=J/4$, $\mu_B = zJ'$ is the chemical potential for holons, and $z$ is the coordination number. 
Applying the mean-field theory such as $\Dt_{ij} = \al{A_{ij}}$ and $\chi_{ij} = \al{B_{ij}}$, 
\begin{align}
	H_{MF} =& -J'\sum_{\al{ij}} (B_{ij}^\dg \chi_{ij} + B_{ij} \chi^*_{ij} + A_{ij}^\dg \Dt_{ij} + A_{ij}\Dt_{ij}^* \nm -|\chi_{ij}|^2  - |\Dt_{ij}|^2) + \mu_B \sum_{i}b_i^\dg b_i \nm - t\chi \sum_{\al{ij}} ( b_i b_j^\dg   + h.c.) 
\end{align}
In path integral formalism, the partition function is given by
\begin{align}
	Z = \int Df D\bar f Db Db^* D\la D\chi D\Dt \exp(-\int_0^\B L_1),
\end{align}
where
\begin{align}
	L_1 =& \sum_i[ \bar f_{i\A} (\pa_\tau -i\la) f_{i\A} + b_i^* (\pa_\tau - i\la+\mu_B ) b_i] \nm + J'\sum_{\al{ij}}( |\chi_{ij}|^2 + |\Dt_{ij}|^2) - J'\chi \sum_{\al{ij}} (\bar f_{i\A} f_{j\A} + h.c.) \nm -  J'\Dt\sum_{\al{ij}}(\bar f_{i\up} \bar f_{j\dw} - \bar f_{i\dw} \bar f_{i\up} + h.c.) \nm  - t\chi \sum_{ij}( b_j^*b_i + h.c.).
\end{align}

For convenience, let us think that the doping $\al{b_i^\dg b_i} = x$ is small, and the phase is uniform resonating valence bond (uRVB), where $\Dt=0$, $\chi,\la \neq 0$. Since the $U(1)$ gauge symmetry is broken, we utilize the $U(1)$ gauge field that obeys 
\begin{align}
	\ab_{ij} \rw \ab_{ij} + \ta_j - \ta_i,
	a_0(i) \rw a_0(i) + \pa_\tau \ta_i,    
\end{align}
where the gauge transform is given by $f_{i\A} \rw f_{i\A}e^{i\ta_i}$ and $b_i \rw b_{i}e^{i\ta_i}$.
Then, we obtain
\begin{align}
	L_1 =& \sum_i[ \bar f_{i\A} (\pa_\tau +\mu_F - i a_0(i)) f_{i\A} \nm + b_i^* (\pa_\tau +\mu_B - ia_0(i)) b_i] \n& - J'\chi \sum_{\al{ij}} (e^{-i\ab_{ij}}B_{ij}^* + e^{i\ab_{ij}} B_{ij} ) \nm - t\chi \sum_{\al{ij}}( e^{i\ab_{ij}}b_i b_j^* + h.c.).
\end{align}
To obtain the effective theory for gauge field, the perturbation theory is introduced to the continuum version of $L_1$, which is
\begin{align}
	L_1 =& \int d^d r ~\bar f_\A(\rb) (\pa_\tau +\mu_F - i a_0(i)) f_\A(\rb) \nm + b^*(\rb) (\pa_\tau +\mu_B - ia_0(i)) b(\rb)] \nm - \f{1}{2m_F} \bar f_\A(\rb)(\grad + i\ab)^2 f_{\A}(\rb) \nm - \f{1}{2m_B} b^*(\rb) (\grad+i\ab)^2 b(\rb),
\end{align}
in $d$-dimension. The coupling between the fermions, bosons and gauge field is 
\begin{align}
	L_{int} = \int d^dr (j_\mu^F + j_\mu^B)\cdot a_\mu.
\end{align}
Integrating over $a_\mu$, we have $j_\mu^F + j_\mu^B = 0$.

Ioffe-Larkin rule can be induced from this condition. Suppose that we have external field $\mathbf{E}$, gauge field $\mathbf{a}$, and associated internal field $\mathbf{e}$. The external field only couples with holons. Then the effective field for fermions and bosons are
\begin{align}
	\mathbf{e}_B = \mathbf{e} + \mathbf{E}, \mathbf{e}_F = \mathbf{e}.
\end{align}
The current density is given by
\begin{align}
	j_F = \ma_F \mathbf{e}_F, j_B = \ma_B \mathbf{e}_B
\end{align}
and $j_F+j_B = 0$ gives
\begin{align}
	\mathbf{e} = -\f{\ma_B}{\ma_F+\ma_B}.
\end{align}
The physical current is
\begin{align}
	j = j_B = -j_F = \f{\ma_F\ma_B}{\ma_F+\ma_B}\mathbf{E}.
\end{align}
Thus, 
\begin{align}
	\ma^{-1} = \ma^{-1}_F + \ma^{-1}_B.
\end{align}
That is, the physical resistivity is the addition of fermionic and bosonic resistivity.
Also, it is note-worthy that in $SU(2)$ gauge theory for $t-J$ model, the Ioffe-Larkin rule is not applicable anymore, since $SU(2)$ gauge field is not directly added to the external fields which is $U(1)$.

\section{ 1D spin chain and its generalization to 2D square lattice }

We begin from the 1D helical spin chain model, which is given by
\begin{align}
	H = -J_1 \sum_i \Sb_i \cdot \Sb_{i+1} + J_2 \sum_i \Sb_i \cdot \Sb_{i+2}, \label{eq:1}
\end{align}
We utilize the Slave-fermion (Schwinger-boson) theory in this model, where $a\equiv f_{i\up}$ and $b\equiv f_{i\dw}$. Then,
\begin{align}
	H =& -\f{J_1}{2}\sum_i \chi_{i,i+1}^\dg \chi_{i,i+1} - \f{J_2}{2}\sum_i \Dt_{i,i+2}^\dg \Dt_{i,i+2} \n& + \sum_i \la_i (n_i - 2S).
\end{align}
where
\begin{align}
	&\chi_{i,i+1} = a_i^\dg a_{i+1} + b_i^\dg b_{i+1}, \n
	&\chi_{i,i+1}^\dg = a_{i+1}^\dg a_{i} + b_{i+1}^\dg b_{i}, \n
	&\Dt_{i,i+2} = a_i b_{i+2} - b_i a_{i+2},\n
	&\Dt_{i,i+2}^\dg = b_{i+2}^\dg a_i^\dg - a_{i+2}^\dg b_i^\dg,\n
	&n_i = a_i^\dg a_i + b_i^\dg b_i,
\end{align}
$\chi_{i,i+1}$ is the nearest-neighbor hopping and $\Dt_{i,i+2}$ is the next-nearest-neighbor spin-singlet coupling of Schwinger-bosons. $\la_i$ is the Lagrange multiplier or chemical potential that impose the constraint on the number of Schwinger-bosons at each site.

The mean-field theory renders  $\al{\chi_{i,i+1}} = \al{\chi_{i,i+1}^\dg} = \chi, \al{\Dt_{i,i+2}} = \al{\Dt_{i,i+2}^\dg} = \Dt$, $\la_i$ = $\la$. The Hamiltonian becomes
\begin{align}
	H_{MF} =& -\f{J_1}{2}\chi \sum_i (\chi_{i,i+1} + \chi_{i,i+1}^\dg) \nm - \f{J_2}{2}\Dt \sum_i (\Dt_{i,i+2} + \Dt_{i,i+2}^\dg) \nm+ \la \sum_i (n_i - 2S) \label{eq:5},
\end{align}
The transform onto $k$-space gives
\begin{align}
	H = \sum_{k} \eta_{\A}^\dg(k) H_{\A\B}(k) \eta_{\B}(k),
\end{align}
where
\begin{align}
	\eta^\dg (k) = \mtx{ a_{k}^\dg & \de b_{-k} }, \eta(k) = \mtx{ a_{k} \\ \de b_{-k}^\dg },
\end{align}
and
\begin{align}
	H(k) = (\la - J_1\chi \cos k)\mathbf{1} + J_2\Dt \sin 2k ~\tau_1.
\end{align}
Here, $\de b_{-k} = -i b_{-k}, \de b^\dg_{-k} = i b_{-k}^\dg$. This can be considered as the bosonic Bogoliubov-de Gennes (BdG) Hamiltonian.

\subsection{Green function}

The Green function is defined as
\begin{align}
	G_{\A\B}(\tau, k) =& -\al{T_\tau \eta_{\A}(\tau) \eta_\B^\dg}.
\end{align}
This can be obtained from the equation of motion of Green functions.
\alg{
	\f{d}{d\tau}G_{\A\B}(\tau,k) =& -\dt(\tau) [\eta_\A,\eta_\B^\dg] + \al{T_\tau \f{d\eta_\A(\tau)}{d\tau}\eta_\B^\dg}.
}
Considering that
\alg{
	&\f{d\eta_\A(\tau)}{d\tau} = [H,\eta_\A(\tau)],
}
where
\alg{
	H =& \sum_k h_{11} a_k^\dg a_k + h_{12} a_k^\dg b_{-k}^\dg + h_{21} b_{-k}a_{k} \nm + h_{22} b_{-k} b_{-k}^\dg,
}
we obtain
\alg{
	~&[H,a_q] = -h_{11} a_q  - h_{12} b_{-q}^\dg,\n
	~&[H,b_{-q}^\dg] = h_{21}a_q + h_{22} b_{-q}^\dg.
}
Thus, the equations of motion are
\alg{
	\f{d}{d\tau}G_{1\B}(\tau,k) =& -\dt(\tau) \dt_{1\B} - h_{11} G_{1\B}(\tau,k) \nm - h_{12} G_{2\B}(\tau,k),\n
	\f{d}{d\tau}G_{2\B}(\tau,k) =& \dt(\tau) \dt_{2\B} + h_{21} G_{1\B}(\tau,k) \nm + h_{22} G_{2\B}(\tau,k).
}
Fourier transform of imaginary time ($d/d\tau \rw -i\omega_n$) gives
\alg{
	-i\w{n} G_{1\B}(i\w{n},k) =&- \dt_{1\B} - h_{11} G_{1\B}(i\w{n},k) \nm- h_{12} G_{2\B}(i\w{n},k),\n
	-i\w{n} G_{2\B}(i\w{n},k) =&  \dt_{2\B} + h_{21} G_{1\B}(i\w{n},k) \nm + h_{22} G_{2\B}(i\w{n},k).
}
In matrix form, this is equivalent to
\alg{
	i\w{n}\tau_3 G(i\w{n},k) = 1 + H_k G(i\w{n},k),
}
where $\tau_i~ (i=0,1,2,3)$ is the Pauli matrix. 
Hence, the Green function is
\alg{
	G(i\w{n},k) = \f{-i\w{n}\tau_3 - (\la - J_1\chi \cos k)1 + J_2 \Dt \sin 2k \tau_1}{\w{n}^2 + \w{k}^2},
}
where $\w{k} = \sqrt{(\la- J_1\chi\cos k)^2 - (J_2 \Dt \sin 2k)^2}$ is the quasi-particle energy in the system. Note that in the ground state, $\w{k}^2 > 0$ for all $k$.

\subsection{Self-consistent ground state}

The inverse Fourier transform gives
\alg{
	\lim_{\tau\rw0^-}G(\tau,k) =& \f{1}{\B}\sum_{i\w{n}} e^{i\w{n}0^+} G(i\w{n},k) \nq -\oint \f{dz}{2\pi i} n(z) e^{z0^+} G(z,k) 
	\nq \f{1}{2} \tau_3 - \f{1+2n(\w{k})}{2\w{k}}[(\la - J_1\chi \cos k)1 \nm - J_2 \Dt \sin 2k \tau_1].
}
According to the definition,
\alg{
	\lim_{\tau\rw0^-}G(\tau,k) = -\mtx{ \al{a_k^\dg a_{k}} & \al{b_{-k}a_k} \\ \al{a_{k}^\dg b_{-k}^\dg} & \al{b_{-k}b_{-k}^\dg} }.
}

\begin{figure}[t]
	\centering
	\includegraphics[width=0.48\columnwidth]{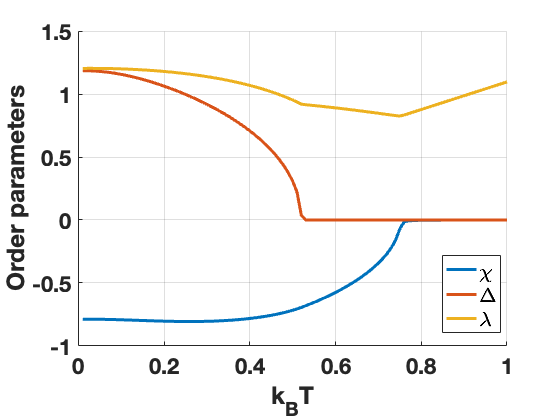}
	\includegraphics[width=0.48\columnwidth]{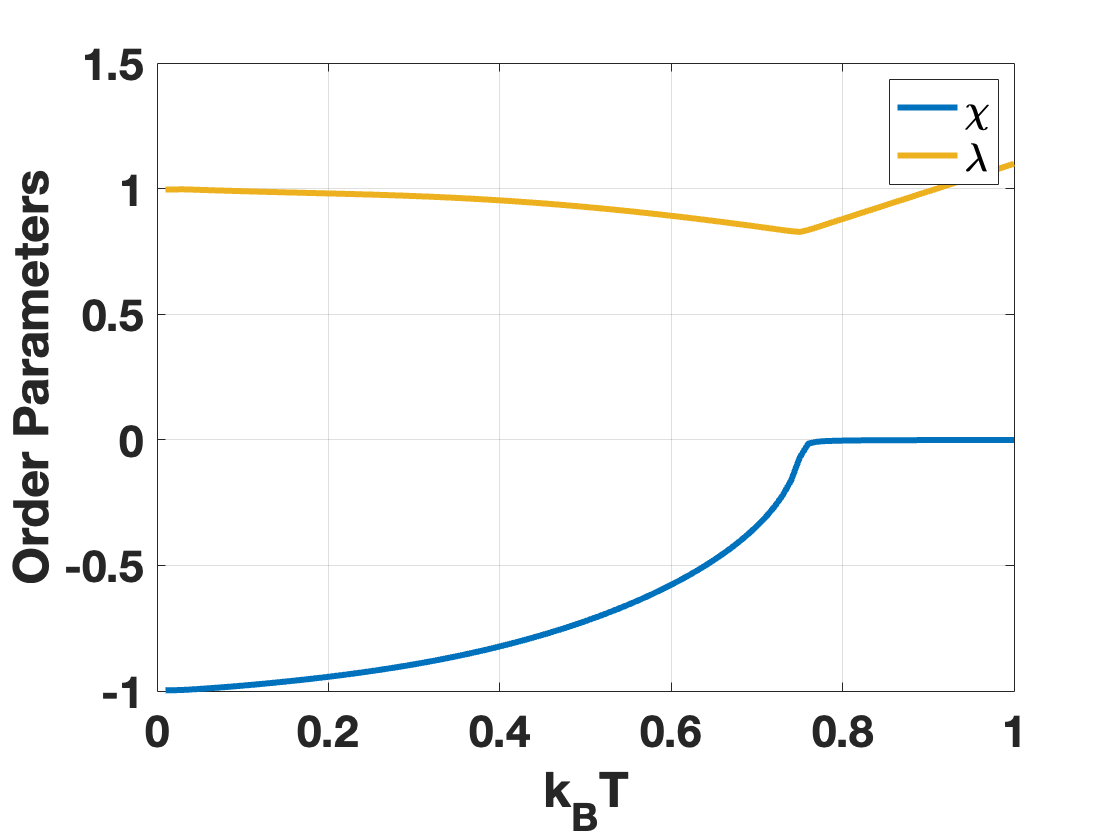}
	\includegraphics[width=.48\columnwidth]{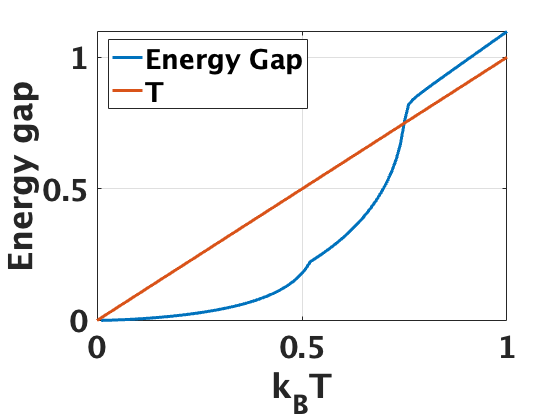}
	\includegraphics[width=.48\columnwidth]{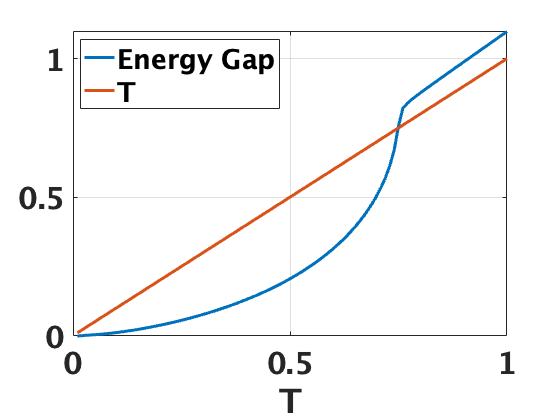}
	\caption{(upper panels) The order parameters in temperatures from self-consistent ground state. The temperatures are in units of $J_1$. (lower panels) The energy gap compared with temperatures. (left panels) $J_2/J_1 = 0.5$, (right panels) $J_2/J_1 = 0.$}
	\label{fig:2-1}
\end{figure}

From this, we obtain the self-consistent equations for order parameters. First, the nearest-neighbor hopping parameter $\chi$ is given by
\begin{align}
	\chi =& \al{\chi_{i,i+1}} = \f{1}{N}\sum_k \al{\ah_k^\dg \ah_k+\bh_{k}^\dg \bh_{k}}e^{ik} 
	\nq \int_{-\pi}^\pi \f{dk}{2\pi}\cos k \f{1+2n(\omega_k)}{\omega_k}(\la - J_1 \chi \cos k).
\end{align}
Next, the next-nearest-neighbor spin-singlet coupling is given by
\begin{align}
	\Dt =& \al{\Dt_{i,i+2}} = \f{1}{N}\sum_k \al{\ah_k \bh_{-k} - \bh_{k}\ah_{-k}} e^{-2ik} 
	\nq  \int_{-\pi}^\pi \f{dk}{2\pi} \f{J_2\Dt \sin^2 2k}{\omega_k} (1+2n(\omega_k)).
\end{align}
At last, the chemical potential can be determined by
\begin{align}
	2S+1 = &\int_{-\pi}^\pi \f{dk}{2\pi} [(\f{\la-J_1 \chi \cos k}{\omega_k} )(2n(\omega_k)+1).
\end{align}

The calculated order parameters at $J_2/J_1 = 0.5$ and $0$ in temperatures are shown in Fig.~\ref{fig:2-1}. At $J_2/J_1 = 0.5$, the ordering temperature for $\chi$ is $T_\chi \approx 0.75$, and that for $\Delta$ is $T_\Dt \approx 0.52$. At $J_2/J_1 = 0$, the ordering temperature for $\chi$ is $T_\chi \approx 0.75$. The energy gap (from zero energy to the lowest spinon dispersion) in temperatures is given also in Fig.~\ref{fig:2-1}.

\subsection{Current-current correlation function}

The current operator from the Hamiltonian is given by
\begin{align}
	J(r_j) =& \f{J_1\chi }{2}i[\chi_{j-1,j} - \chi_{j-1,j}^\dg].
\end{align}
The Fourier transform of current operator is
\alg{
	J(q) = -J_1\chi\sum_{k}[a_{k_-}^\dg a_{k_+} + b_{k_-}^\dg b_{k_+} ]\sin k,
}
where $k_\pm = k\pm q/2$.

The current-current (auto-)correlation function is given by
\alg{
	\Pi(q;\tau) =& -\al{T_\tau J(q,\tau)J(-q)}
}
After some algebra, we arrive at
\alg{
	\Pi(q,\tau) =&  -(J_1\chi)^2\sum_{k} \sin^2 k \nm\times \Tr[G(-\tau,k_-)\tau_3 G(\tau,k_+)\tau_3].
}

The Fourier transform of correlation function is
\alg{
	\Pi(q,i\w{n}) =& -\f{(J_1\chi)^2}{\B} \sum_{k} \sin^2k \sum_{i\w{l}} \nm\times  \Tr[G(i\w{l},k_-)\tau_3 G(i\w{l}+i\w{n},k_+)\tau_3] .
}
The tedious algebra gives 
\alg{
	\Pi(q,i\w{n}) = (J_1\chi)^2 \sum_k \sin^2k~ \f{f}{g},
}
where
\alg{
	g =& \w{k_-}\w{k_+}((\w{k_-}-\w{k_+})^2 + \w{n}^2)\nm\times ((\w{k_-}+\w{k_+})^2 + \w{n}^2)
}.
and 
\alg{
	f =& (\w{k_-} + \w{k_+})(\w{k_-} \w{k_+} - \xi) ((\w{k_-}-\w{k_+})^2 + \w{n}^2) 
	\nm + 2n(\w{k_-}) \w{k_+} [(\w{k_-}^2 - \w{k_+}^2)(\xi+\w{k_-}^2) \nm+ \w{n}^2(-\xi +\w{k_-}^2)]
	+ 2n(\w{k_+}) \w{k_-} \nm\times [(\w{k_+}^2 - \w{k_-}^2)(\xi+\w{k_+}^2)  + \w{n}^2(-\xi + \w{k_+}^2)].
}
The retarded correlation function is just obtained by analytic continuation $i\w{n} \rw \w{}+i\eta$. $\eta$ is related to the lifetime of spinons.

When $J_2/J_1= 0$, the system only has the ferromagnetic exchange, and the correlation function is reduced to
\alg{
	\Pi^R(q,\w{}) = 2(J_1\chi)^2 \int_k \sin^2 k \f{\Dt \w{k} \Dt n_k}{\Dt \w{k}^2 -(\w{}+i\eta)^2}, \label{eq:50}
}
where $\Dt \w{k} = \w{k_-} -\w{k_+}$, $\Dt n_k = n(\w{k_-}) - n(\w{k_+})$, $n$ is Bose-Einstein distribution, and $\int_k = d^dk/(2\pi)^d$. $d=1$ in this case.

The inductance can be obtained as follows. First, the real conductivity can be obtained from the imaginary part of correlation function, 
\alg{
	\Re \ma(\w{}) = -\f{\Im \Pi^R}{\w{}}.
}
Then, the imaginary conductivity can be computed by the Kramers-Kronig relation.
\alg{
	\Im \ma(\w{}) = -\f{1}{\pi} \int d\w{}' \f{\Re \ma(\w{}')}{\w{}'-\w{}},
}
Then, the resistivity is given by $\rho(\w{}) = \ma(\w{})^{-1}$, and the inductivity is found as
\alg{
	\mathcal{L}(\w{}) = -\f{\Im \rho(\w{})}{\w{}}.
}
and the inductance is $L = \mathcal{L}l/A$, where $l$ is the length and $A$ is the area of the system. We let $l=1, A=1$ for all cases.

\subsection{Drude-type inductance}

In Eq.~\ref{eq:50}, suppose $\eta \rw 0$. From the fact,
\alg{
	\lim_{\eta\rw 0} \f{1}{\w{}+i\eta} = P \f{1}{\w{}} -i\pi \dt(\w{}),
}
we obtain
\alg{
	\Pi^R(q,\w{}) =& -2(J_1\chi)^2 \int_k \sin^2 k \f{\Dt \w{k} \Dt n_k}{2\w{}}[\f{2\w{}}{\w{}^2-\Dt\w{k}^2}\nm -i\pi \dt( \f{\w{}^2-\Dt\w{k}^2}{2\w{}})].
}
Using another fact
\alg{
	\dt(g(x)) = \sum_i \f{\dt(x-x_i)}{|g'(x_i)|},
}
we have
\alg{
	\Re \ma(q,\w{}) =& 2\pi(J_1\chi)^2 \int_k \sin^2 k (-\Dt \w{k} \Dt n_k)\nm\times[\f{\dt(\w{}+\Dt\w{k})+\dt(\w{}-\Dt\w{k})}{2\w{}^2}].
}
The Kramers-Kronig relation leads to
\alg{
	\Im \ma(q,\w{}) =& -(J_1\chi)^2 \int_k \sin^2 k (-\f{\Dt n_k}{\Dt \w{k}}) \nm\times\f{2\w{}}{\Dt\w{k}^2 - \w{}^2}.
}
After some brief algebra, the total conductivity is given by
\alg{
	\ma(q,\w{}) =& (J_1\chi)^2 \int_k \sin^2 k (-\f{\Dt n_k}{\Dt \w{k}})  [\pi\dt(\f{\w{}^2 -\Dt \w{k}^2}{2\w{}}) \nm + i\f{2\w{}}{\w{}^2-\Dt\w{k}^2}]
}
When $q\rw 0$,
\alg{
	\ma(0,\w{}) =& 2(J_1\chi)^2 \int_k \sin^2 k (-\f{\partial n(\w{k})}{\partial \w{k}} ) (\pi \dt(\w{}) + \f{i}{\w{}}) \nq C(\pi \dt(\w{}) + \f{i}{\w{}}).
}
Here,
\alg{
	C = 2(J_1\chi)^2 \int_k \sin^2k (-\f{\partial n(\w{k})}{\partial \w{k}} ) > 0.
} Thus, the inductance is given by
\alg{
	L = \f{1}{C},
}
which is just constant in frequencies. Since $C$ is proportional to $\chi^2$, $L$ reaches the highest peak near $T_\chi$ where $\chi$ becomes small. It is worth noting that although we revive $\eta$,
\alg{
	\ma(0,\w{}) = C(\f{1}{-i\w{}+\eta}), 
}
but the inductance is still $L=1/C$.

\subsection{Dirty system}

\begin{figure}
	\centering
	\includegraphics[width=0.48\columnwidth]{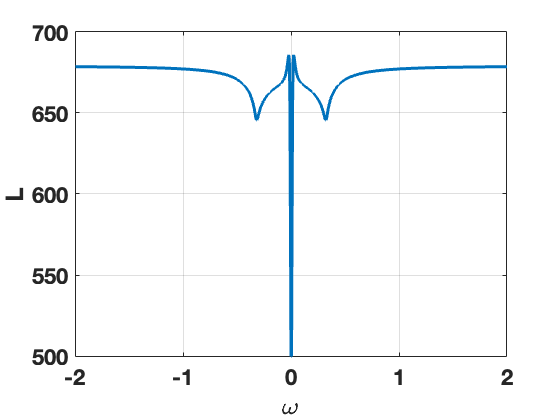}
	\includegraphics[width=0.48\columnwidth]{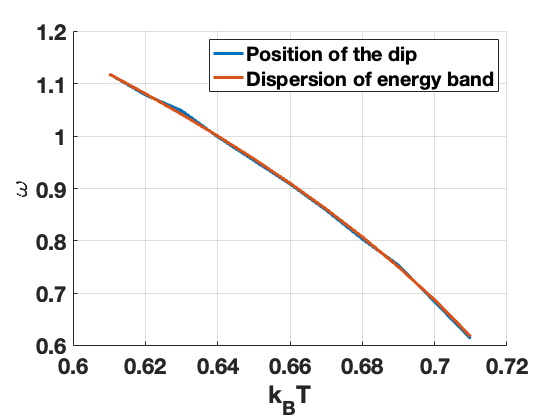}
	\caption{(left panel) The inductance in 1D ferromagnetic spin with impurities near $T_\chi$. (right panel) Comparison of the position of dip at $\w{}\neq 0$ and dispersion of energy band in temperatures.}
	\label{fig:2-3}
\end{figure}

When the system is dirty enough by impurities and disorders, the system is well described by the the Mattis-Bardeen scheme.
\alg{
	\Pi_d^R(0,\omega) = \sum_q \f{2\A}{q^2 + \A^2}\Pi^R(q,\w{})
}
The correlation function in temperature and $\omega$ is computed by setting $\eta = 1\times10^{-2}$ and $\A=1/200$. The associated inductance near $T_\chi$ is shown in the left panel of Fig.~\ref{fig:2-3}. Notably, different from the Drude-type, the two distinct dip structures at $\w{}=0$ and $\w{}\neq 0$ appear. Both dips are due to the intraband transition $\Dt \w{k}$ in Eq.~\ref{eq:50}. The dip at $\w{}=0$ are attributed to $q \sim 0$, and the dip at $\w{}\neq 0$ are attributed to $q = q_T$, where $q_T$ is the momentum distance difference between lowest and highest energy $k$-points. When we define the dispersion of energy band as the difference of highest and lowest energies, one can notice that the dip position and the dispersion of energy band are the same as shown in the right panel of Fig.~\ref{fig:2-3}.

\subsection{Generalization to 2D square lattice}

The results we obtained here is readily generalized to the higher dimensions with a single band, for instance, a ferromagnetic 2D square lattice. The current-current correlation function for the ferromagnetic 2D square lattice has the same form as Eq.~\ref{eq:50}, but $\sin^2 k \rw \sin^2 k_i$ ($i=x,y$) and $\w{k}^2 = \la - J_1\chi (\cos k_x +\cos k_y)$. Therefore, by the same procedure in the previous part, the inductance of the spinons in a 2D square lattice is constant in frequencies as well.

\subsection{The unit of inductance}

Here, the unit we use is $J_1 = 1 \sim 1$ meV, $\hbar = e = 1$, and the lattice constant $a = 1 \sim 1~\AA$. Thus, $k_BT = 1 \sim 11.6$ K, $\w{} = 1 \sim 242$ MHz, $\tau = 1 \sim 4.13 \times 10^{-9} ~s$, $\rho = 1 \sim 25.8~ k\Omega \cdot \AA = 258 ~\mu \Omega \cdot cm$. $\mathcal{L} = 1 \sim ~ 1.07 ~pH\cdot cm$. At $T=T_\chi$, $\mathcal{L}\sim 10^3 \approx 1 ~nH\cdot cm$. If the system is micro-sized, $L \sim 10^3 \approx 10~\mu H$, which is observable in experiments.

We suppose that the holon has the average inductance of metal. The typical value of resistivity and relaxation time is given by $\rho_0  \sim 10 ~\mu\Omega \cdot cm$, $\tau \sim 1\times 10^{-9}~s$. $\mathcal{L} = \rho_0\tau = 10~fH\cdot cm$, which is negligible compared to that of spinons.

\section{ 2D Honeycomb lattice }

Let us consider the Hamiltonian only with the nearest-neighbor ferromagnetic exchanges. 
\alg{
	H =& - J \sum_{\al{ab}} \Sb_a \cdot \Sb_b
	\nq - J \sum_{ij} \Sb_{ijA} \cdot \Sb_{ijB} + \Sb_{ijA} \cdot \Sb_{i+1j B} \nm + \Sb_{ijA} \cdot \Sb_{ij+1B}.
}
The unit cell is in Fig.~\ref{fig:3}. The red dots denote A sublattice, and the blue dots denote B sublattice.

\begin{figure}[t]
	\centering
	\includegraphics[width=\columnwidth]{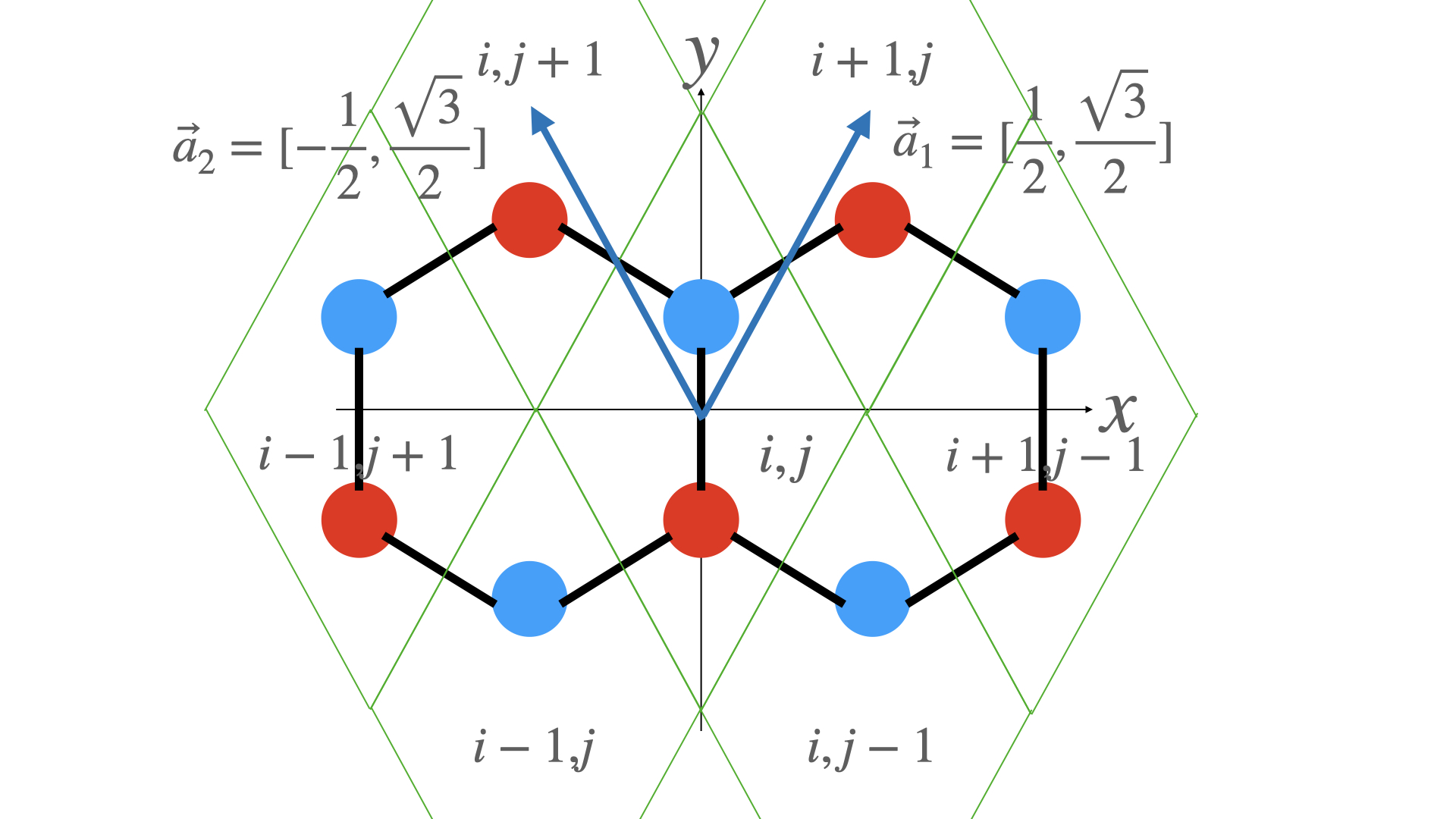}
	\caption{The unit cells of honeycomb lattice. The red dots are A sublattices, and the blue dots are B sublattices.}
	\label{fig:3}
\end{figure}

\subsection{Mean-Field theory}

The Slave-fermion (Schwinger-boson) theory for ferromagnetic honeycomb lattice is
\alg{
	H =& -J \sum_{ij} [\chi_{ijA,ijB}^\dg \chi_{ijA,ijB} + \chi_{i+1jA,ijB}^\dg \chi_{i+1jA,ijB} \nm + \chi_{ij+1A,ijB}^\dg \chi_{ij+1A,ijB}] \nm + \sum_{ij} \sum_{\xi=A}^B \la_{ij\xi}(n_{ij\xi}-2S),
}
where $\xi$ is the sublattice index and
\alg{
	\chi_{ij\xi,i'j'\xi'} = a_{i'j'\xi'}^\dg a_{ij\xi} + b_{i'j'\xi'}^\dg b_{ij\xi}, \n
	\chi_{ij\xi,i'j'\xi'}^\dg = a_{ij\xi}^\dg a_{i'j'\xi'} + b_{ij\xi}^\dg b_{i'j'\xi'}.
}

Let us apply the mean field theory
\alg{
	\chi = \al{\chi_{ij\xi,i'j'\xi'}} = \al{\chi_{ij\xi,i'j'\xi'}^\dg}, \la = \la_{ij\xi}.
}
Then,
\alg{
	H_{MF} =& -J \chi \sum_{ij} [ \chi_{ijA,ijB}^\dg + \chi_{ijA,ijB} + \chi_{i+1jA,ijB}^\dg \nm + \chi_{i+1jA,ijB} + \chi_{ij+1A,ijB}^\dg + \chi_{ij+1A,ijB} ]\nm + \la \sum_{ij} \sum_{\xi=A}^B (n_{ij\xi}-2S).
}
The Fourier transform gives
\alg{
	H_{MF} =& -J\chi \sum_{k} [(a_{kA}^\dg a_{kB} + b_{kA}^\dg b_{kB})(1+ e^{-ik_1} + e^{-ik_2}) \nm+ (a_{kB}^\dg a_{kA} + b_{kB}^\dg b_{kA})(1+ e^{ik_1} + e^{ik_2})] \nm + \la \sum_{k,\xi} (a_{k\xi}^\dg a_{k\xi} + b_{k\xi}^\dg b_{k\xi} - 2S).
}
In matrix form
\alg{
	H_{MF} = V_{\A}(k)^\dg H_{\A\B}(k) V_{\B}(k),
}
where
\alg{
	&V = [a_{kA}, a_{kB}, b_{kA}, b_{kB}]^T,\nm
	V^\dg = [a_{kA}^\dg,a_{kB}^\dg,b_{kB}^\dg,b_{kB}^\dg],
}
and
\alg{
	H(k) =& \la \tau_0 \ma_0 - J \chi( (1+\cos k_1 + \cos k_2)\tau_0\ma_1 \nm + (\sin k_1 + \sin k_2) \tau_0\ma_2).
}
Here, $\tau_i$ are the Pauli matrices representing the spin degrees of freedom, and $\ma_i$ are that representing the sublattice degrees of freedom. The energy is given by
\alg{
	E = \la \pm \w{k},
}
where $\w{k}^2 = (J\chi)^2 (3 + 2\cos k_1 + 2 \cos k_2 + 2\cos(k_1-k_2))$. The energy bands are doubly degenerate. Also, the system has the band crossings at $K$ points in the Brillouin zone.

Since this is the block-diagonal and all blocks are the same, we utilize only one block.
\alg{
	H(k) =& \la \ma_0 - J \chi( (1+\cos k_1 + \cos k_2) \ma_1 \nm + (\sin k_1 + \sin k_2) \ma_2).
}

\subsection{Green function and Self-consistent Equations}

Here the Green function is simply given by
\alg{
	G(i\w{n},k) =& (i\w{n} - H)^{-1} \nq \f{1}{(i\w{n}-\la)^2 -  \w{k}^2 }[(i\w{n}-\la) \ma_0 \nm - J\chi[(1+\cos k_1 + \cos k_2)\ma_1 \nm + (\sin k_1 + \sin k_2) \ma_2 ].
}
We should find
\alg{
	G_{\A\B}(0^-,k) = -\al{ V_\B^\dg(k) V_\A(k)}
}
The inverse Fourier transform is
\alg{
	G(0^-,k) =& \f{1}{\B}\sum_{i\w{n}} e^{i\w{n}0^+} G(i\w{n},k).
}
When $n_{tot} = n(\la+\omega_k) + n(\la-\omega_k)$, $\Dn = n(\la-\w{k}) - n(\la+\w{k})$, the Green function becomes
\alg{
	G(0^-,k) =& -\frac{1}{2}\{n_{tot}\ma_0 + \f{-\Dn}{\w{k}}(-J\chi)\nm\times [(1+\cos k_1 + \cos k_2)\ma_1 \nm + (\sin k_1+\sin k_2)\ma_2 ]\}.
}

\begin{figure}[t]
	\centering
	\includegraphics[width=0.48\columnwidth]{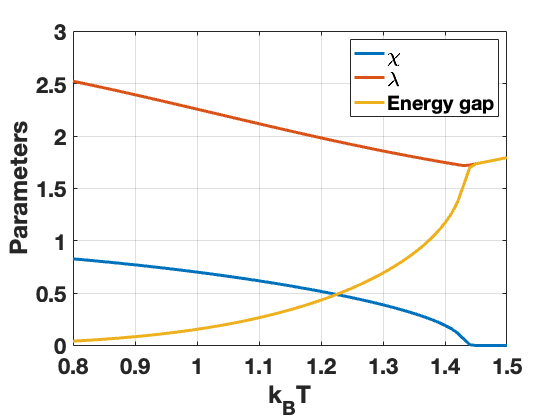}
	\includegraphics[width=0.48\columnwidth]{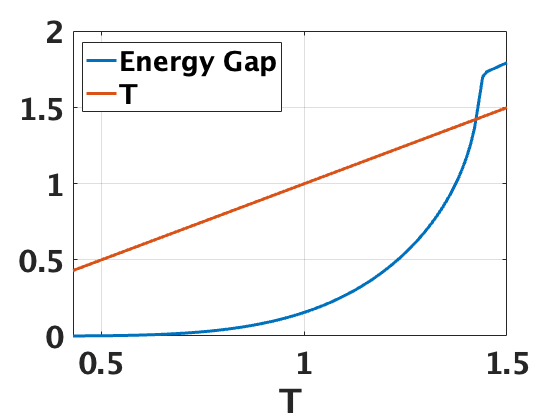}
	\caption{(left panel) The order parameters in temperatures and (right panel) the energy gap compared with temperatures from self-consistent ground state in the ferromagnetic 2D honeycomb SFMFT. }
	\label{fig:2}
\end{figure}

We can find the self-consistent equations from here.
\alg{
	\chi =& -J\chi \int \f{d^2k}{(2\pi)^2} [\f{-\Dn}{\w{k}}]  (1+\cos k_1 + \cos k_2).
}
Also, $\la$ is determined by
\alg{
	2S =& \int \f{d^2 k}{(2\pi)^2} n_{tot}.
}
The ground state acquired self-consistently in Fig.~\ref{fig:2} with $J=1$. The transition temperature is $T_{\chi} \sim 1.45 $.

\subsection{The current-current correlation function}

The current operator is given by
\alg{
	J_{1ij} = i(-J\chi)[\chi_{ijA,i-1jB}^\dg - \chi_{ijA,i-1jB}],\n
	J_{2ij} = i(-J\chi)[\chi_{ijA,ij-1B}^\dg - \chi_{ijA,ij-1B}].
}
The Fourier transform gives
\alg{
	J_1(\qb) =&  \f{i(-J\chi)}{\sqrt N}\sum_{k} a_{k_-A}^\dg a_{k_+B} e^{-ik_{1}} - a_{k_-B}^\dg a_{k_+ A} e^{ik_{1}} \nm (a\rw b), \n
	J_2(\qb) =&  \f{i(-J\chi)}{\sqrt N}\sum_{k} a_{k_-A}^\dg a_{k_+B} e^{-ik_{2}}- a_{k_-B}^\dg a_{k_+ A} e^{ik_{2}} \nm (a\rw b).
}

From the current operator, we can acquire the current-current correlation function. Defining $\int_k = \int \f{d^2k}{(2\pi)^2}$,
\alg{
	\Pi_{11}(\tau,q) =& -\al{T_\tau J_1(\tau,q) J_1(-q)}, 
}
When we define
\alg{
	P_1(k) = \sin k_1 \tau_1 - \cos k_1 \tau_2,
}
the Fourier transform gives
\alg{
	\Pi_{11}(i\w{n},q) =& -\f{(J\chi)^2}{\B} \int_k \sum_{i\w{l}} \Tr G(i\w{l},k_-) P_1(k)\nm\times  G(i\w{l}+i\w{n},k_+) P_1(k).
}

\begin{figure}[t]
	\centering
	\includegraphics[width=\columnwidth]{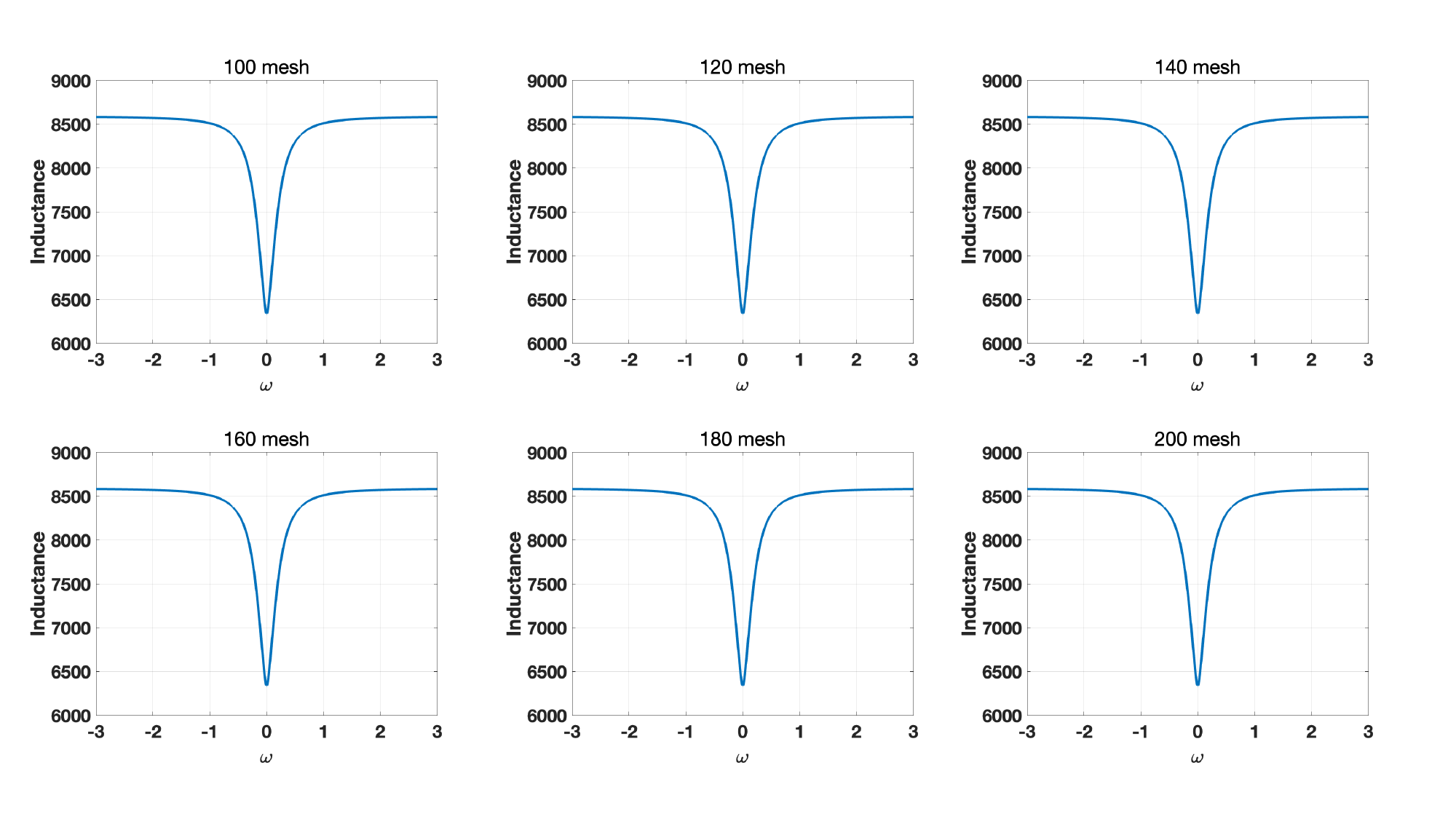}
	\caption{The computed inductance in 2D honeycomb lattice near $T_{\chi}$. $J=1$, $\eta=10^{-1}$ are fixed, but $k$-mesh sizes are changed from $100\times100$ to $200\times200$. }
	\label{fig:5}
\end{figure}

\begin{figure}[t]
	\centering
	\includegraphics[width=\columnwidth]{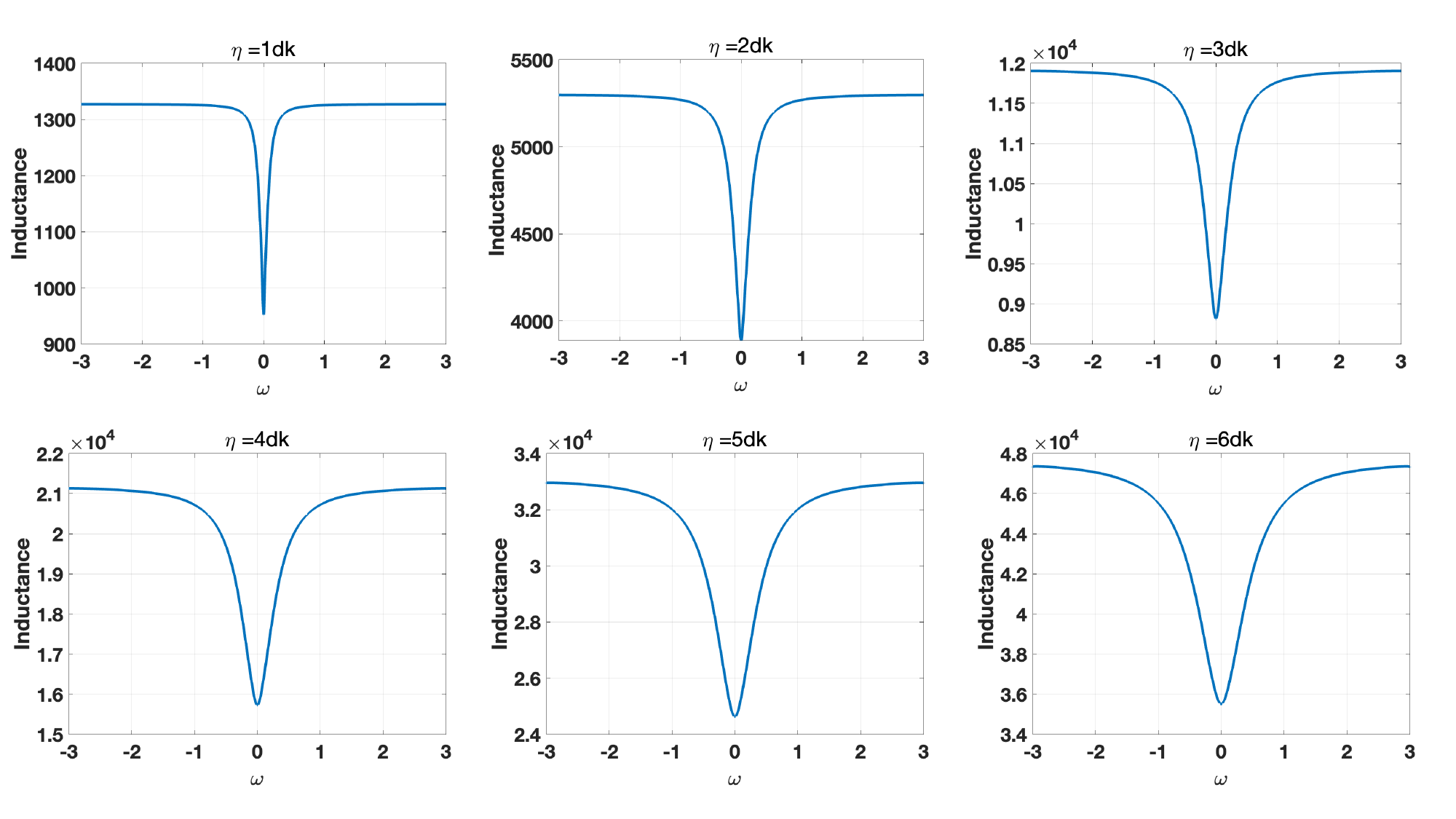}
	\caption{The computed inductance in 2D honeycomb lattice near $T_{\chi}$. $J=1$, $k$-mesh sizes ($160\times 160$) are fixed, but $\eta$ changes from $1 dk$ to $6 dk$.}
	\label{fig:6}
\end{figure}

We denote the function 
\alg{
	F_1(\kb,\qb) =& 1+\cos 2k_1 + \cos 2(k_1-k_2) + \cos k_{-1} + \cos k_{+1} \nm 
	+ \cos(k_{+1}-k_{+2})  + \cos(k_{-1} - k_{2-}) \nm+ \cos (2k_1-k_{+2}) + \cos(2k_1-k_{-2}).
}
Then, the long algebraic process ends at
\alg{
	\Pi_{11}(i\w{n},\qb) =& 
	(J\chi)^2 \int_k \f{1}{[(\w{-}-\w{+})^2+\w{n}^2][(\w{-}+\w{+})^2+\w{n}^2]}\nm
	\times\{ \Dt n_- [\w{-}(\w{+}^2-\w{-}^2-\w{n}^2) \nm + \f{(J\chi)^2 F_1}{\w{-}}(\w{-}^2-\w{+}^2-\w{n}^2) ]
	\nm+\Dt n_+ [\w{+}(\w{-}^2-\w{+}^2-\w{n}^2) \nm+ \f{(J\chi)^2 F_1}{\w{+}}(\w{+}^2-\w{-}^2-\w{n}^2)]\}.
}
where $\Dt n_\pm \equiv n(\la-\w{\pm}) - n(\la + \w{\pm})$.
The retarded correlation function is obtained by analytic continuation $i\w{n} \rw \w{}+i\eta$.

When $\qb\rightarrow 0$ with $\Dt n_k = n(\la-\w{k}) - n(\la+\w{k})$, the correlation function is now
\alg{
	\Pi(\w{},0) = 2(J\chi)^2 \int_k \f{\Dt n_k}{\w{k}}\f{(J\chi)^2 F_1(\kb,0) + \w{k}^2}{(\w{}+i\eta)^2 - 4\w{k}^2}. \label{eq:89}
}
We show the numerical computation of inductance from Eq.~\ref{eq:89} in Fig.~\ref{fig:5} and Fig.~\ref{fig:6}. The dip structure appears near $\w{}=0$ for every case. In Fig.~\ref{fig:5}, we change the $k$-mesh size from $100\times100$ to $200\times200$ and show that the result is independent of $k$-mesh size. In Fig.~\ref{fig:6}, we change $\eta$, which is related to the lifetime of spinons,c from $1 dk$ to $6 dk$ where $dk$ is the length of a side of $k$-mesh plaquette. The dip structure depends on $\eta$. Remarkably, when $\eta$ increases, the width of dip is enlarged, the depth of dip is decreased (in percentage), and the saturated value of the inductance is increased.

To explain the behavior of inductance, let us suppose $\eta\rightarrow 0$, we can write this as
\alg{
	\Pi(\w{},0) =& 4(J\chi)^4 \int_k \f{\Dt n_k}{\w{k}}\f{g_1^2}{2\w{}}[\f{2\w{}}{\w{}^2-4\w{k}^2} \nm - i\pi \dt(\f{\w{}^2-4\w{k}^2}{2\w{}})]
}
where $g_1(\kb) = 1+\cos k_1 + \cos(k_1-k_2) $.
The real conductivity is
\alg{
	\Re\ma(\w{},0) =& \pi (J\chi)^4 \int_k \f{\Dt n_k}{\w{k}}\f{g_1^2}{2\w{k}^2}[ \dt(\w{}-2\w{k})\nm+ \dt(\w{}+2\w{k})].
}
The imaginary conductivity is obtained by the Kramers-Kronig relation.
\alg{
	\Im \ma(\w{},0) =& (J\chi)^4 \int_k \f{\Dt n_k}{\w{k}}\f{g_1^2}{2\w{k}^2}[ \f{2\w{}}{\w{}^2-4\w{k}^2}].
}
Thus, the conductivity is
\alg{
	\ma(\w{},0) =&  2(J\chi)^4 \int_k \f{\Dt n_k}{\w{k}}\f{g_1^2}{2\w{k}^2}[ \pi \dt(\f{\w{}^2-4\w{k}^2}{2\w{}}) + i \f{2\w{}}{\w{}^2-4\w{k}^2}]  \nq
	\lim_{\eta \rw 0} 2(J\chi)^4 \int_k \f{\Dt n_k}{\w{k}}\f{g_1^2}{\w{k}^2} \nm\times[\f{i(\w{}+i\eta)}{(\w{}+i\eta)^2-(2\w{k})^2}] .
}
The factor $2$ comes from the spin degeneracy. Notably, the conductivity comes from the interband transition, since the energy difference between energy bands at $k$-point is $2\w{k}$. It is worthy noting that the integrand shows resonance at $\w{} = 0$ if there is a band crossing point, i.e., $\w{k}=0$. Since the honeycomb lattice has the band crossings at $K$ points $(\w{k} = 0)$, conductivity is greatly enhanced near $\w{}=0$, and the resistivity and inductance is reduced near $\w{}=0$. Furthermore, according to the structure of the integrand, the dip structure is related to $\eta$ as well. Hence, this explains the sharp dip structure appears in the inductance at $\w{}=0$.

From this result, we can also infer that when the system is bipartite and the energy bands are gapped throughout the whole Brillouin zone ($\w{k}\neq 0$), the system would not show the dip structure at $\w{}=0$. This is because the integrand in the above equation never shows resonance near $\w{}=0$ anymore.  

\subsection{Dirty system}

\begin{figure}
	\centering
	\includegraphics[width=0.7\columnwidth]{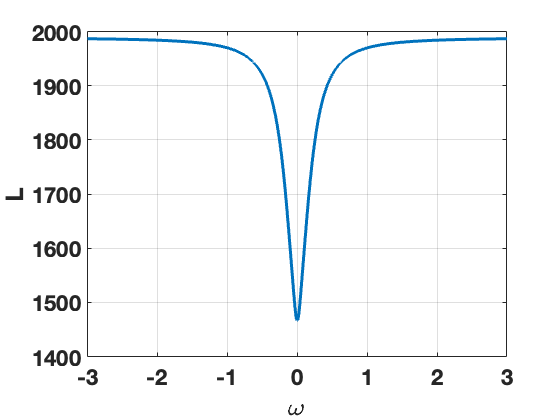}
	\caption{The inductance in 2D honeycomb lattice with impurities near $T_{\chi}$.}
	\label{fig:7}
\end{figure}

For the dirty system, we compute
\alg{
	\Pi^d(\w{}) =& \int \f{d^2q}{(2\pi)^2} \f{4\A}{q(q^2+\A^2)}\Pi^R(\w{},\qb).
}
We let the $k$-mesh size $160\times 160$ and $\eta = 10^{-1}$, $\alpha = 1/50$, and numerically compute the inductance near $T_{\chi}$ in Fig.~\ref{fig:7}. This shares the same structural properties as Figs.~\ref{fig:5} and \ref{fig:6}.

\section{ 2D Kagomé lattice}

\begin{figure}[t]
	\centering
	\includegraphics[width=\columnwidth]{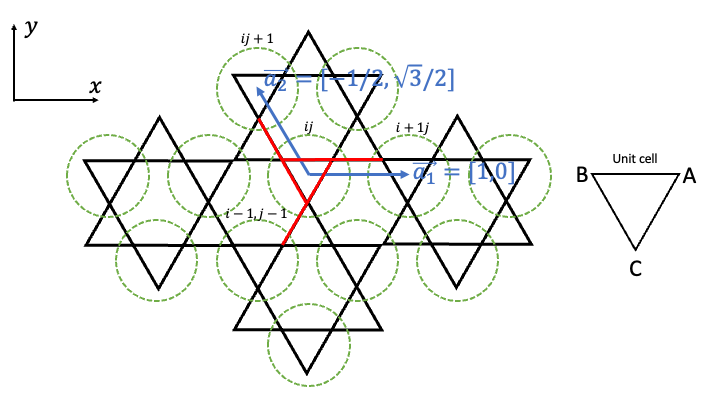}
	\caption{The unit cells of Kagomé lattice.}
	\label{fig:8}
\end{figure}

Let us consider the Hamiltonian only with the nearest-neighbor ferromagnetic exchanges.
\alg{
	H =& -J \sum_{\al{ab}} \Sb_a \cdot \Sb_b
	\nq -J \sum_{ij} [\Sb_{ijA} \cdot \Sb_{ijB} + \Sb_{ijB} \cdot \Sb_{ijC} + \Sb_{ijC} \cdot \Sb_{ijA} \nm+ \Sb_{ijA} \cdot \Sb_{i+1j B} + \Sb_{ijB} \cdot \Sb_{ij+1C} \nm + \Sb_{ijC} \cdot \Sb_{i-1j-1A} ].
}
The upper line is the intra cell, and the lower line is the inter cell contributions. The unit cell is shown in Fig.~\ref{fig:8}.

We use the following notations. $k_i = \kb \cdot \vec a_i$, $k_i' = \kb' \cdot \vec a_i$, $\vec a_1 = (1,0), \vec a_2 = (-1/2,\sqrt3/2)$, $\vec b_1 = 2\pi (1,1/\sqrt{3})$, $\vec b_2 = 2\pi(0,2/\sqrt3)$. $\kb = (k_1 \vec b_1 + k_2 \vec b_2)/(2\pi)$. Within the Brillouin zone we take $k_1,k_2 \in [-\pi,\pi]$. The Brillouin zone area is $8\pi^2/\sqrt3$.

\subsection{Mean-Field theory}

The Slave-fermion (Schwinger-boson) theory gives
\alg{
	H =& -J \sum_{ij} [\chi_{ijA,ijB}^\dg \chi_{ijA,ijB} + \chi_{ijB,ijC}^\dg \chi_{ijB,ijC} \nm + \chi_{ijC,ijA}^\dg \chi_{ijC,ijA} + \chi_{ijA,i+1jB}^\dg \chi_{ijA,i+1jB} \nm + \chi_{ijB,ij+1C}^\dg \chi_{ijB,ij+1C} + \chi_{ijC,i-1j-1A}^\dg \chi_{ijC,i-1j-1A} \nm+ \sum_{ij} \sum_{\xi=A}^C \la_{ij\xi}(n_{ij\xi}-2S),
}
where $\f{J}{2} \rw J$ is redefined, $\xi$ is the sublattice index, and
\alg{
	\chi_{ij\xi,i'j'\xi'} =& a_{i'j'\xi'}^\dg a_{ij\xi} + b_{i'j'\xi'}^\dg b_{ij\xi}, \n
	\chi_{ij\xi,i'j'\xi'}^\dg =& a_{ij\xi}^\dg a_{i'j'\xi'} + b_{ij\xi}^\dg b_{i'j'\xi'}.
}

Let us apply the mean field theory
\alg{
	\chi = \al{\chi_{ij\xi,i'j'\xi'}} = \al{\chi_{ij\xi,i'j'\xi'}^\dg}, \la = \la_{ij\xi}.
}
Then,
\alg{
	H_{MF} =& -J \chi \sum_{ij} [ \chi_{ijA,ijB}^\dg + \chi_{ijA,ijB} + \chi_{ijB,ijC}^\dg + \chi_{ijB,ijC} \nm
	+ \chi_{ijC,ijA}^\dg + \chi_{ijC,ijA} + \chi_{ijA,i+1jB}^\dg \nm + \chi_{ijA,i+1jB} + \chi_{ijB,ij+1C}^\dg + \chi_{ijB,ij+1C} \nm + \chi_{ijC,i-1j-1A}^\dg + \chi_{ijC,i-1j-1A}] \nm + \la \sum_{ij} \sum_{\xi=A}^B (n_{ij\xi}-2S), \label{eq11}
}
The Fourier transform gives
\alg{
	H_{MF} =& - J\chi \sum_{k}[ (a_{kB}^\dg a_{kA} + b_{kB}^\dg b_{kA} ) (1+e^{-ik_1}) \nm + (a_{kC}^\dg a_{kB} + b_{kC}^\dg b_{kB} ) (1+e^{-ik_2}) \nm + (a_{kA}^\dg a_{kC} + b_{kA}^\dg b_{kC} ) (1+e^{i(k_1+k_2)}) + h.c. ] \nm + \la \sum_{k,\xi} (a_{k\xi}^\dg a_{k\xi} +  b_{k\xi}^\dg b_{-k\xi} - 2S).
}
In matrix form
\alg{
	H_{MF} = V_{\A}(k)^\dg H_{\A\B}(k) V_{\B}(k),
}
where
\alg{
	&V(k) = [a_{kA}, a_{kB}, a_{kC}, b_{kA}, b_{kB}, b_{kC}]^T,\nm
	V^\dg(k) = [a_{kA}^\dg, a_{kB}^\dg, a_{kC}^\dg, b_{kA}^\dg, b_{kB}^\dg, b_{kC}^\dg].
}
and
\alg{
	H(k) = [\la I_{3\times3} - J\chi P ]\otimes \ma_0,
}
where 
\alg{
	P = \mtx{
		0  & 1+e^{ik_1}  &  1+e^{i(k_1+k_2)} \\
		1+e^{-ik_1}  & 0  & 1+e^{ik_2}  \\
		1+e^{-i(k_1+k_2)}  &  1+e^{-ik_2} & 0    
	}.
}
Here, $\ma_i$ are that representing the sublattice degrees of freedom. The energy is given by
\alg{
	E = \la + \w{i,k},
}
where
\alg{
	\w{1,k} =& -J\chi(1+\ep_k), \n
	\w{2,k} =& -J\chi(1-\ep_k), \n
	\w{3,k} =& 2J\chi.
}
Here, $\ep_k^2 = (3 + 2\cos k_1 + 2 \cos k_2 + 2\cos(k_1+k_2)) = (1+\cos k_1 + \cos k_2)^2 + (\sin k_1 - \sin k_2)^2$. The energy bands are doubly degenerate. There are fourfold band crossings between $\w{1,k}$ and $\w{2,k}$ at $(k_1,k_2) = \pm 2\pi (-1/3,-1/3)$, $\pm 2\pi (-2/3,1/3)$, $\pm 2\pi(1/3,-2/3)$ which are $K$-points in the Brillouin zone (i.e., $\ep_k =0$). Also, there are band crossings between $\w{2,k}$ and $\w{3,k}$ at $\kb = 0 (\Gamma)$. In the calculation, we consider one block since spin up and down are degenerate.

\subsection{Green function and self-consistent equation}

The Green function is 
\alg{
	G(i\w{n},k) = (i\w{n}-H)^{-1}.
}
and what we should find here is
\alg{
	G(0^-,k) =& \f{1}{\B}\sum_{i\w{n}} e^{i\w{n}0^+} G(i\w{n},k).
}

\begin{figure}
	\centering
	\includegraphics[width=0.48\columnwidth]{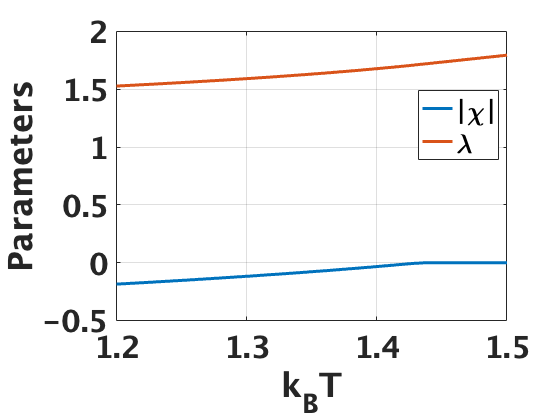}
	\includegraphics[width=0.48\columnwidth]{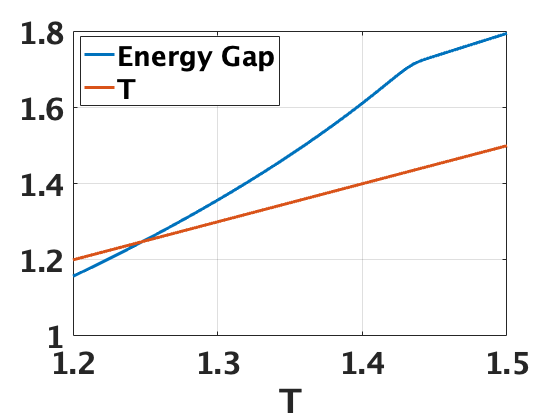}
	\caption{(left panel) The order parameters in temperatures and (right panel) the energy gap compared with temperatures from self-consistent ground state in the ferromagnetic 2D Kagomé spin model.}
	\label{fig:9}
\end{figure}

Because of the complexity, we here acquire the Green function in Lehmann representation. Using the identity $\sum_{a=1}^3 \ket{a,k}\bra{a,k} = I$ where $\ket{a,k}$ is the eigenstate of $H(k)$ with eigenvalue $\la + \w{a,k}$, one finds
\alg{
	G(i\w{n},k) =& \sum_{a,b} \ket{a,k}\bra{a,k} (i\w{n}-H)^{-1} \ket{b,k}\bra{b,k}, \n
	=&\sum_a \f{1}{i\w{n}-\la - \w{a,k}} \ket{a,k}\bra{a,k}.
}
Then,
\alg{
	G(0^-,k) =& - \sum_a \ket{a,k}\bra{a,k} [\int \f{dz}{2\pi i} \f{n(z)}{z-\la-\w{a,k}} ] \nq
	- \sum_a \ket{a,k}\bra{a,k} (n(\la + \w{a,k})).
}

We can find the self-consistent equations from here. ($\int_k = \int d^2k/(2\pi)^2$.)
\alg{
	\chi =& -\f{1}{3}\int_k [G_{12}(0^-,k) + G_{21}(0^-,k) +G_{23}(0^-,k) \nm + G_{32}(0^-,k) + G_{31}(0^-,k) + G_{13}(0^-,k) ].
}
Also,
\alg{
	2S =& -\f{2}{3} \int_k \Tr[G(0^-,k)].
}
The transition occurs at $T_{\chi} = 1.43$.

\subsection{The current-current correlation function}

The current operator in this system is
\alg{
	J_1(i,j) =& -iJ\chi ( \chi_{i-1jA,ijB}^\dg -  \chi_{i-1jA,ijB}), \n
	J_2(i,j) =& -iJ\chi (\chi_{ij-1B,ijC}^\dg- \chi_{ij-1B,ijC}), \n
	J_3(i,j) =& -iJ\chi (\chi_{i+1j+1C,ijA}^\dg  -\chi_{i+1j+1C,ijA} ),
}
The Fourier transform gives
\alg{
	J_1(\qb) =& -i\f{J\chi}{\sqrt{N}} \sum_{k}[a_{k_-A}^\dg a_{k_+B} e^{ik_1} - a_{k_-B}^\dg a_{k_+ A} e^{-ik_1}]. \n
	J_2(\qb) =& -i\f{J\chi}{\sqrt{N}} \sum_{k}[a_{k_-B}^\dg a_{k_+C} e^{ik_2} - a_{k_-C}^\dg a_{k_+ B} e^{-ik_2} ]. \n
	J_3(\qb) =& -i\f{J\chi}{\sqrt{N}} \sum_{k}[a_{k_-C}^\dg a_{k_+A} e^{-i(k_1+k_2)} \nm - a_{k_-A}^\dg a_{k_+ C} e^{i(k_1+k_2)} ].
}
Here, $\kb_\pm = \kb \pm \qb/2$.

\begin{figure}
	\centering
	\includegraphics[width=0.7\columnwidth]{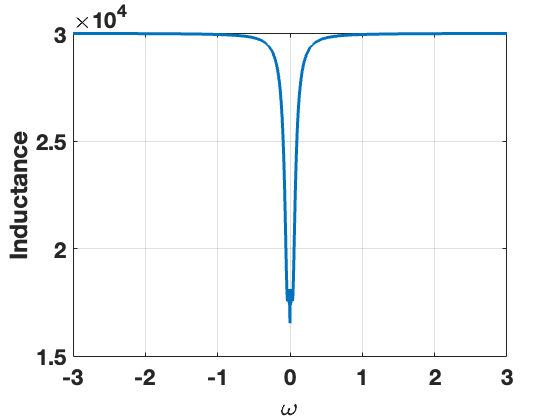}
	\caption{The inductance in 2D Kagomé lattice near $T_{\chi}$.}
	\label{fig:10}
\end{figure}

Let us find the following correlation function.
\alg{
	\Pi_{11}(\tau,\qb) = - \al{T_\tau J_1(\tau,\qb) J_1(-\qb)}.
}
Then,
\alg{
	\Pi_{11}(\tau,\qb) =& -(J\chi)^2 \int_k \Tr G(-\tau,k_-) P_1(k) G(\tau,k_+) \nm\times  P_1(k),
}
where
\alg{
	P_1(k) = -i\mtx{ 
		0  & e^{ik_1}  &  0 \\
		-e^{-ik_1}  & 0  & 0  \\
		0  &  0 & 0
	}.
}
The Fourier Transform gives
\alg{
	\Pi_{11}(i\w{n},\qb) =&
	-\f{(J\chi)^2}{\B}  \int_k \sum_{i\w{l}} \Tr G(i\w{l},k_-) P_1(k) \nm \times G(i(\w{l}+\w{n}),k_+) P_1(k), 
}

We denote $U$ and $U'$ the unitary matrix diagonalizing $H(k)$ at $k=k_-$ and $k=k_+$ each, $\Dn_{ab} = n(\la+\w{b,k_+}) - n(\la+\w{a,k_-}) , \Dt \w{ab} = \w{b,k_+} - \w{a,k_-}$.
Thus,
\alg{
	\Pi_{11}(i\w{n},\qb) =& 
	(J\chi)^2 \int_k \sum_{ab} \f{-\Dn_{ab} \Dw_{ab}}{(i\w{n}-\Dw_{ab})(i\w{n}+\Dw_{ab})}\nm\times [U_{2,a} (U^\dg)_{a,2} U_{1,b}' (U^\dg)'_{b,1} \nm + U_{1,a} (U^\dg)_{a,1} U_{2,b}' (U^\dg)'_{b,2} \nm - U_{1,a} (U^\dg)_{a,2} U_{1,b}' (U^\dg)'_{b,2}e^{-2ik_1} \nm - U_{2,a} (U^\dg)_{a,1} U_{2,b}' (U^\dg)'_{b,1}e^{2ik_1}  ]. \label{eq:119}
}
The retarded form is obtained from the analytic continuation $i\w{n} \rw \w{}+i\eta$. When $q = 0$, $\Dw_{ab}$ represents the interband transition of spinons. It is note-worthy that the correlation function and the conductivity shows the resonance when $\w{}=0$ since there are the band crossing points $\Dw_{ab} = 0$. Thus, it is expected to have the dip structure near $\w{}=0$ as well as the 2D honeycomb lattice. The expectation is confirmed as shown in the numerical computation result in Fig.~\ref{fig:10} ($J=1, \eta = 10^{-1}, T \sim T_{\chi}$). As well as the case of 2D honeycomb, the computation is independent of $k$-mesh size, and the dip width and depth depend strongly on the parameter $\eta$. Furthermore, the inductance in the dirty system also shows the same structure.

This result could be generalized into the arbitrary number of energy bands. Although the current-current correlation function changes its detailed form, 
but the denominator is always represented as
\alg{
	(\w{}+i\eta - \Dw_{ab})(\w{}+i\eta + \Dw_{ab}).
}
in Lehmann representation. Since $\Dw_{ab}$ represents the interband transition, the band crossing point ($\Dw_{ab}=0$) gives the resonance near $\w{}=0$, leading to the sharp dip near $\w{}=0$. 
However, if there is no band crossing, there will be no resonance near $\w{}=0$, the sharp dip structure will vanish. 
When the system is dirty, now $\Dw_{ab}$ also contains the intraband transitions, the sharp dip structure at $\w{}=0$ will appear.
We illustrate the statement by an example next section.

\section{ 1D spin ladder chain}

\begin{figure}[ht]
	\centering
	\includegraphics[width=.6\columnwidth]{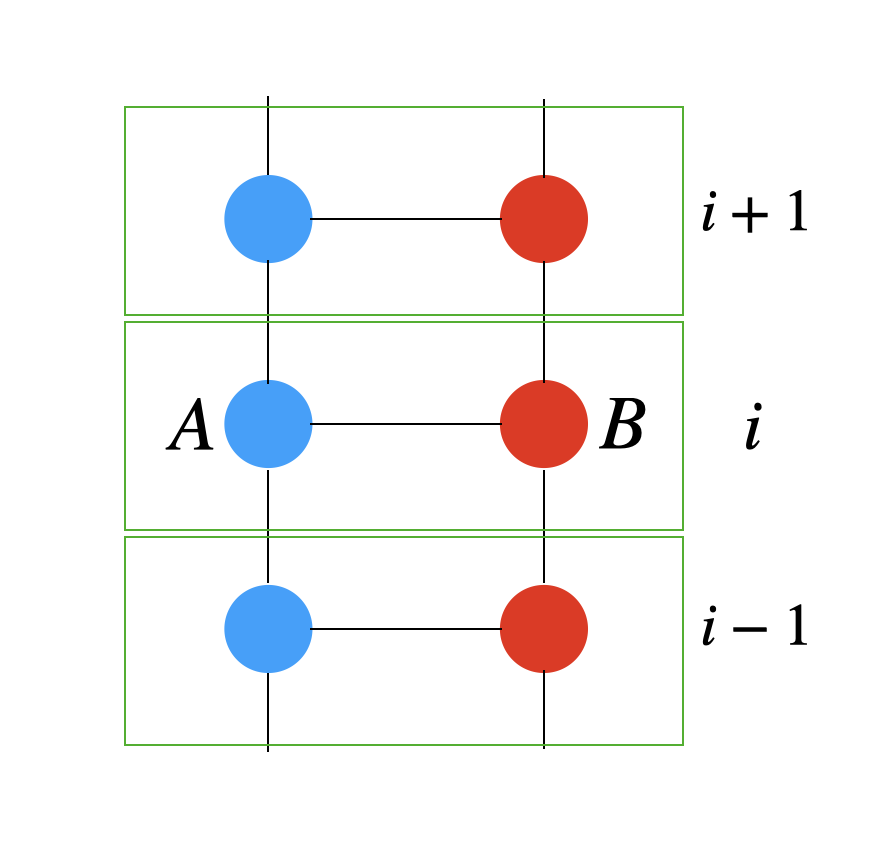}
	\caption{The unit cell of 1D spin ladder chain.}
	\label{fig:11}
\end{figure}

For illustrating the inductance in a gapped multiband system, we take 1D spin ladder chain model.
\alg{
	H =& - J\sum_{\al{ab}} \Sb_a \cdot \Sb_b \nq
	- J \sum_{i} [\Sb_{iA}\cdot \Sb_{iB} + \Sb_{iA} \cdot \Sb_{i+1A} + \Sb_{iB} \cdot \Sb_{i+1B}].
}
The structure of 1D spin ladder chain is given in Fig.~\ref{fig:11}.

\subsection{Mean-Field theory}

The Slave-fermion (Schwinger-boson) theory for ferromagnetic ladder model is 
\alg{
	H =& -J \sum_{i} [\chi_{iA,iB}^\dg \chi_{iA,iB} + \chi_{iA,i+1A}^\dg \chi_{iA,i+1A} \nm+ \chi_{iB,i+1B}^\dg \chi_{iB,i+1B}] +  \sum_{i,\A} \la_{i\A} [n_{i\A} - 1]\nm
	-h\sum_{i\A} [n_{iA} - n_{iB}].
}
Here, $\A = A,B$, and $h$ gives additional imbalance between $A$ and $B$ and enlarges the energy gap. The mean-field theory gives,
\alg{
	H =& -J\chi \sum_{i} [\chi_{iA,iB}^\dg +\chi_{iA,iB} + \chi_{iA,i+1A}^\dg \chi_{iA,i+1A} \nm+ \chi_{iB,i+1B}^\dg \chi_{iB,i+1B}] + \la \sum_{i,\A} [n_{i\A} - 1] \nm - h \sum_{i} [n_{iA} - n_{iB}].
}
In $k$-space, there are two copies of
\alg{
	H(k) = (\la-2J\chi\cos k)\tau_0 - J\chi \tau_1 - h\tau_3.
}
We only care $\up$ spin and double the result. The energy is
\alg{
	E_\pm(k) = \la - 2J\chi \cos k \pm b,
}
where $b = \sqrt{(J\chi)^2 + h^2}$. The energy is gapped about $2b$ throughout the whole Brillouin zone.
We represent the Green function in Lehmann representation.
\alg{
	G(k,i\w{n}) =& (i\w{n}-H)^{-1} \nq
	\sum_a \f{\ket{a}\bra{a}}{i\w{n}-E_a(k)}
}
where $a=\pm$.

\subsection{Current-current correlation function}

The current operator is
\alg{
	J(i) =& (-iJ\chi)[\chi_{iA,i+1A}^\dg - \chi_{iA,i+1A} \nm+ \chi_{iB,i+1B}^\dg - \chi_{iB,i+1B} .]
}
By Fourier transform, one could get
\alg{
	J(q) =& \f{J\chi}{N} \sum_k \sin k[u_{k_-,A}^\dg u_{k_+ A} + u_{k_-,B}^\dg u_{k_+ B} ]
}
where $u=f_\up$, $k_\pm = k\pm q/2$.

The current-current correlation function is given by
\alg{
	\Pi(\qb,i\w{n}) =& -\f{(J\chi)^2}{\B}\int_k \sin^2 k\sum_{i\w{l}} \nm\times\Tr[G(i\w{n}+i\w{l},k_-) G(i\w{l},k_+)]
}
After some algebra with the Green function above and analytic continuation, one would find out that
\alg{
	\Pi(q,\w{}) = 2(J\chi)^2 \int_k \sin^2 k \sum_a \f{-\Dt n_a \Dt E_a}{(\w{}+i\eta)^2- \Dt E_a^2},
}
where $\Dt n_a = n_B(E_{a}(k_-)) - n_B(E_{a}(k_+))$ and $\Dt E_a = E_a(k_-) - E_a(k_+)$. Here the factor $2$ comes from inclusion of the $\dw$ spin part. When $\eta \rw 0$,
\alg{
	\Pi(q,\w{}) =& 2(J\chi)^2 \int_k \sin^2 k \sum_a (-\f{\Dt n_a}{ \Dt E_a}) \f{\Dt E_a^2}{2\w{}} \nm\times[\f{2\w{}}{\w{}^2- \Dt E_a^2}-i\pi \dt(\f{\w{}^2 - \Dt E_a^2}{2\w{}}) ].
}

The conductivity is given by
\alg{
	\ma(q,\w{}) =& 2(J\chi)^2 \int_k \sin^2 k \sum_a (-\f{\Dt n_a}{\Dt E_a}) [\pi \dt(\f{\w{}^2 - \Dt E_a^2}{\w{}})\nm +i\f{\w{}}{\w{}^2 - \Dt E_a^2} ].
}
It is note-worthy that the correlation function and conductivity has the same form as that of 1D spin chain, so we anticipate the constant inductance.
When we let $q\rw 0$ and revive $\eta$,
\alg{
	\ma(\w{}) =& \lim_{\eta\rw0} 2(J\chi)^2 i\int_k \sum_a (-\f{\Dt n_a}{\Dt E_a})\f{1}{\w{}+i\eta} 
	\nq \lim_{\eta\rw 0} \f{iC}{\w{}+i\eta}.
}
Here, $C>0$ depends only on the temperature. Thus, the inductance $L=1/C$ is constant in frequencies just as the anticipation. 

%


\end{document}